%% file: main.tex
\documentclass[sigconf]{acmart} %anonymous

\usepackage{booktabs} % For formal tables
\usepackage{multirow}
\usepackage{subcaption}
\usepackage{xparse}
\usepackage{bm}
\usepackage{bbm}
\usepackage{amsmath}

\usepackage{dsfont}
\usepackage{soul}
\usepackage{tabularx}
\usepackage{enumitem}
\usepackage{balance}

\usepackage{algpseudocode}
\usepackage{algorithm}

\usepackage{graphicx}

\NewDocumentCommand{\Log}{o}{\IfNoValueTF{#1}{}{{}^{#1}\!}\log}
\DeclareMathOperator*{\argmax}{argmax}

\copyrightyear{2019}
\acmYear{2019} 
\setcopyright{iw3c2w3}
\acmConference[WWW '19]{Proceedings of the 2019 World Wide Web Conference}{May 13--17, 2019}{San Francisco, CA, USA}
\acmBooktitle{Proceedings of the 2019 World Wide Web Conference (WWW '19), May 13--17, 2019, San Francisco, CA, USA}
\acmPrice{}
\acmDOI{10.1145/3308558.3313632}
\acmISBN{978-1-4503-6674-8/19/05}

\usepackage{balance}

\newcommand{\nop}[1]{}
\newcommand{\fromJ}[1]{\textcolor{blue}{#1}}
\newcommand{\fromJJ}[1]{\textcolor{brown}{From Jay: #1}}
\newcommand{\fromZ}[1]{\textcolor{orange}{From Ziyu: #1}}

\newcommand{\add}[1]{\textcolor{purple}{#1}}

\begin{document}
\title{CoaCor: Code Annotation for Code Retrieval with Reinforcement Learning}

\author{Ziyu Yao}
\affiliation{%
  \institution{The Ohio State University}
%   \streetaddress{Dreese Labs}
%   \city{Columbus}
%   \postcode{43210}
}
\email{yao.470@osu.edu}

\author{Jayavardhan Reddy Peddamail}
% \orcid{0000-0003-4233-1222}
\affiliation{%
  \institution{The Ohio State University}
%   \streetaddress{Dreese Labs}
%   \city{Columbus}
%   \postcode{43210}
}
\email{peddamail.1@osu.edu}

\author{Huan Sun}
% \begin{comment}
% \authornote{The secretary disavows any knowledge of this author's actions.}
% \end{comment}
\affiliation{%
  \institution{The Ohio State University}
%   \streetaddress{Dreese Labs}
%   \city{Columbus}
%   \postcode{43210}
}
\email{sun.397@osu.edu}

\begin{abstract}
To accelerate software development, much research has been performed to help people understand and reuse the huge amount of available code resources. Two important tasks have been widely studied: \textit{code retrieval}, which aims to retrieve code snippets relevant to a given natural language query from a code base, and \textit{code annotation}, where the goal is to annotate a code snippet with a natural language description. Despite their advancement in recent years, the two tasks are mostly explored separately. In this work, we investigate a novel perspective of \textit{\underline{Co}de \underline{a}nnotation for \underline{Co}de \underline{r}etrieval} (hence called ``CoaCor''), where a code annotation model is trained to generate a natural language annotation that can represent the semantic meaning of a given code snippet and can be leveraged by a code retrieval model to better distinguish relevant code snippets from others. To this end, we propose an effective framework based on reinforcement learning, which explicitly encourages the code annotation model to generate annotations that can be used for the retrieval task. Through extensive experiments, we show that code annotations generated by our framework are much more detailed and more useful for code retrieval, and they can further improve the performance of existing code retrieval models significantly.\footnote{Code available at \url{https://github.com/LittleYUYU/CoaCor}.}
\end{abstract}

\ccsdesc[500]{Information systems~Novelty in information retrieval}
\ccsdesc[500]{Information systems~Summarization}
\ccsdesc[500]{Software and its engineering}
%%\ccsdesc[500]{Information systems~Learning to rank}
\ccsdesc[300]{Computing methodologies~Reinforcement learning}
\ccsdesc[300]{Computing methodologies~Markov decision processes}
\ccsdesc[100]{Computing methodologies~Neural networks}

\keywords{Code Annotation; Code Retrieval; Reinforcement Learning}

\maketitle

\input{introduction}

\input{preliminaries}
\input{frameworkoverview}
\input{model}
\input{exp}

\input{relatedwork}
\input{conclusion}

\bibliographystyle{ACM-Reference-Format}
\balance
\bibliography{references}

\end{document}

%% file: introduction.tex
\section{Introduction}
\label{sec:intro}
%\textbf{FLOW:}
%\nop{Background about code search, code annotation, and their importance: A lot of scenarios with NL and code snippets, great resources that need to be utilized, two important tasks.}

Software engineering plays an important role in modern society. Almost every aspect of human life, including health care, education, transportation and web security, {depends} on reliable software \cite{allamanis2018survey}. Unfortunately, developing and maintaining large code bases are very costly. Understanding and reusing billions of lines of code in online open-source repositories can significantly speed up the software development process. Towards that, \textit{code retrieval} (CR) and \textit{code annotation} (CA) are two important tasks that have been widely studied in the past few years \cite{allamanis2018survey, gu2018deep, iyer2016summarizing, xing2018DeepCom, wan2018improving}, where the former aims to retrieve relevant code snippets based on a natural language (NL) query while the latter is to generate natural language descriptions to describe what a given code snippet does. 

Most existing work \cite{gu2018deep, xing2018DeepCom, wan2018improving, allamanis2016convolutional, jiang2017automatically, wang2018comment} study either code annotation or code retrieval individually\nop{code annotation and code retrieval separately}. Earlier approaches for code retrieval drew inspiration from the information retrieval field \cite{haiduc2013automatic, lu2015query, hill2014nl, keivanloo2014spotting} and suffered from {surface form} \nop{keyword}mismatches between natural language queries and code snippets \cite{biggerstaff1994program, mcmillan2012exemplar}.
% \nop{Various code retrieval models have been proposed, most of them drawing inspiration from information retrieval (IR) techniques \cite{haiduc2013automatic, lu2015query,hill2014nl, chan2012searching, brandt2010example, lv2015codehow, mcmillan2011portfolio, Vinayakarao:2017:AIS:3018661.3018691}. Earlier IR-based approaches  suffered from a mismatch between natural language queries and source code snippets \cite{biggerstaff1994program, mcmillan2012exemplar,}.}\nop{IR based approaches try to find match by using lexical tokens or language structure similarities, both of which might not exist between a code and a query. An effective code search approaches requires developing a higher-level semantic mapping between code and natural language queries.} 
More recently, advanced deep learning approaches\nop{, which embed natural language queries or {and} code snippets into a high-dimensional vector space{ and learn a high-level semantic mapping between them},} have been successfully applied to both code retrieval and code annotation \cite{allamanis2018survey, gu2018deep, allamanis2016convolutional, xing2018DeepCom, iyer2016summarizing, wang2018comment, jiang2017automatically, loyola2017neural, ijcai2018-314, wan2018improving, chen2018neural}.
For example, {the code retrieval model proposed by} \citet{gu2018deep} utilized two deep neural networks to learn the vector representation of a natural language query and that of a code snippet respectively, and adopted cosine similarity to measure their matching degree. For code annotation, \citet{iyer2016summarizing} and \citet{xing2018DeepCom} utilized encoder-decoder models with an attention mechanism to generate an NL annotation for a code snippet. {They aim to generate annotations similar to the human-provided ones, and therefore trained the models using the {standard} maximum likelihood estimation (MLE) objective.}
For the same purpose, \citet{wan2018improving} trained the code annotation model in a reinforcement learning {(RL)} framework with reward being the BLEU score \cite{papineni2002bleu}, which measures n-gram matching precision between the currently generated annotation and the human-provided one. 

\begin{figure*}[ht!]
\includegraphics[width=0.95\textwidth, height=8.5cm]{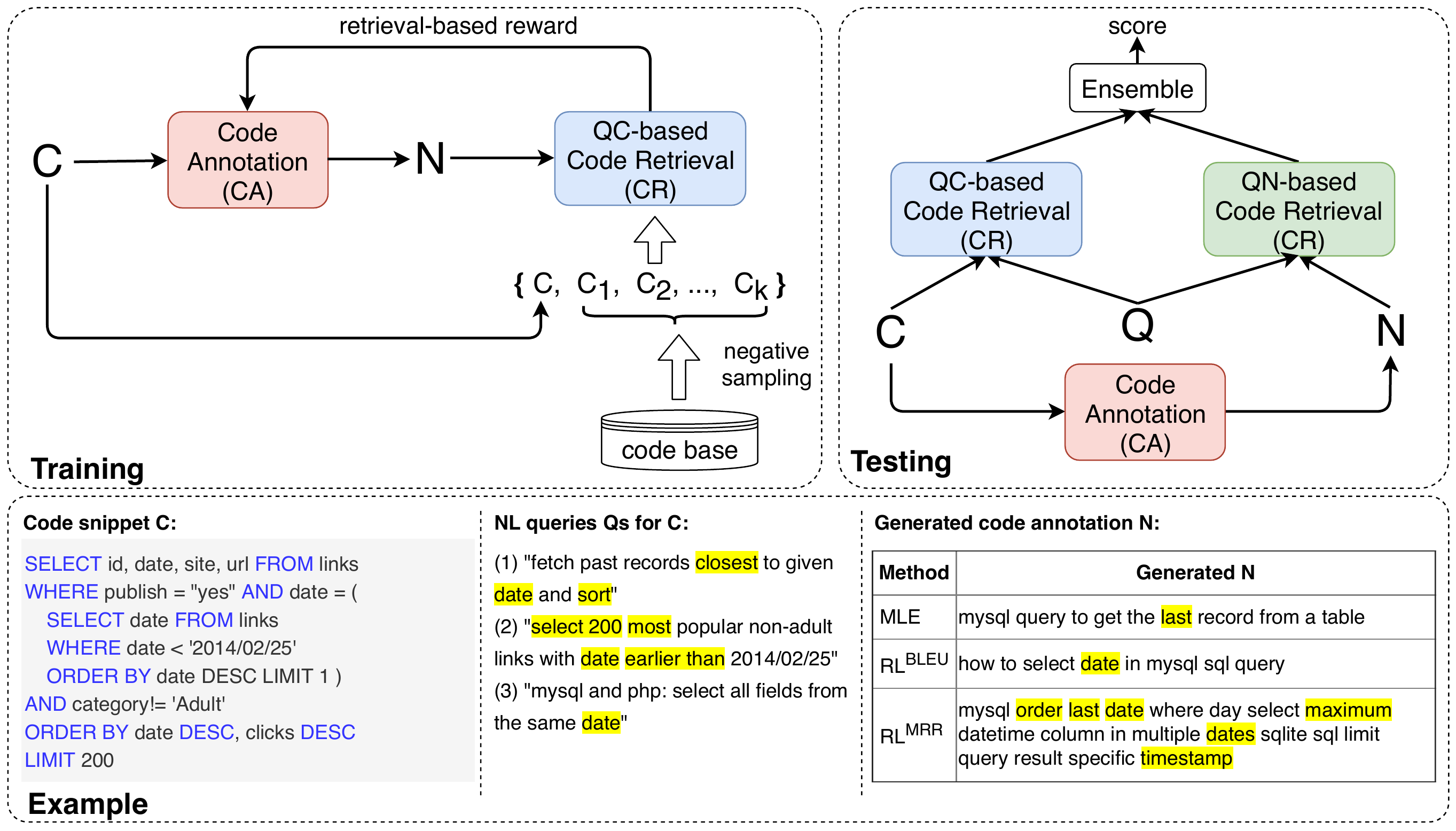}
\caption{Our CoaCor framework: (1) \textit{Training phase}. A code annotation model is trained via reinforcement learning to maximize retrieval-based rewards\nop{a ranking metric based reward or simply ranking rewards?} given by {a QC-based code retrieval model (\nop{\st{i.e\.,}}pre-trained using <NL \ul{q}uery, \ul{c}ode snippet> pairs).} (2) \textit{Testing phase}. {Each code snippet is first annotated by the trained CA model. For the code retrieval task, given query Q, a code snippet gets two scores - one {matching Q with its code content and the other matching Q with its code annotation {N}}, and is ranked by a simple ensemble strategy\textsuperscript{+}.}\nop{\footnote{We simply use a weighted combination of the two scores, and other ensemble strategies can also apply here.}\fromJ{I did not find the footnote 1?}} (3) \textit{Example}. {We show an example of a code snippet and its associated multiple NL queries in our dataset}. The code annotation generated by our framework (denoted as RL\textsuperscript{MRR}) {is much more detailed with many keywords \nop{mentioned}\textit{semantically aligned} with Qs}, {when} compared with CA models trained via {MLE or RL with BLEU rewards (RL\textsuperscript{BLEU}).} \small\textsuperscript{+} We simply use a weighted combination of the two scores, and other ensemble strategies can also apply here.}
\label{fig:actor_critic_framework}
\end{figure*}

In this work, we explore a novel perspective - code annotation \textit{for} code retrieval (CoaCor), which is to generate an NL annotation for a code snippet so that the generated annotation \textit{can be used for\nop{help} code retrieval} (i.e.,\textit{ can represent the semantic meaning of a code snippet and distinguish it from others w.r.t. a given NL query in the code retrieval task}). As exemplified by \cite{Vinayakarao:2017:AIS:3018661.3018691}, such an annotation can be taken as the representation of the corresponding code snippet, based on which {the aforementioned lexical mismatch issue in a naive keyword-based search engine can be alleviated.}\nop{a naive keyword-based search engine can alleviate the aforementioned lexical mismatch issue while maintaining the search efficiency}
A similar idea of improving retrieval by adding extra annotations to items is also explored in document retrieval \cite{scholer2004query} and image search \cite{wiki:google_img_labeler}. However, most of them rely on humans to provide the annotations.\nop{ or, human-provided annotations.}
% \fromH{Do we need to empirically show this keyword-based search?}\fromZ{a. is it expected to surpass DCS? b. better QN model to surpass DCS? (as Q and N are in the same modal, probably a benefit here?)}
Intuitively, \nop{this}{our} perspective can be interpreted as one type of \textit{machine-machine collaboration}: On the one hand, the NL annotation generated by the code annotation model can serve as a second view of a code snippet (in addition to its programming content)\nop{, which can} {and} {can} be utilized to match with an NL query in code retrieval. 
On the other hand, with a goal to facilitate code retrieval, the code annotation model can be stimulated to produce rich and {detailed} annotations.
% \nop{remove? , which improves its generation quality can say this in discussion or future work: Our motivation is that we hope the generated code annotations are useful for other tasks, of which we choose code retrieval as an example. }
Unlike existing work \cite{iyer2016summarizing, xing2018DeepCom, wan2018improving, ijcai2018-314}, our goal is \textit{not} to generate an NL annotation as close as possible to a human-provided one; hence, {the MLE objective or the BLEU score as rewards for code annotation will not fit our setting}. {Instead, we design a {novel} rewarding mechanism in an RL framework, which guides the code annotation model directly based on how effectively the currently generated code annotation distinguishes\nop{distinguishes or retrieves?}\nop{helps retrieve} the code snippet from a candidate set.} 
%\fromH{Unlike existing work, we aim to generate NL annotations for a code snippet so that the generated annotations can help distinguish the code snippet from others in the context of code retrieval.} 

{Leveraging collaborations and interactions among machine learning models to improve the task performance has been explored in other scenarios. \citet{goodfellow2014generative} proposed the Generative Adversatial Nets (GANs), where a generative model produces difficult examples to fool a discriminative model and the latter is further trained to conquer the challenge. \citet{he2016dual} proposed another framework called \textit{dual learning}, which jointly learns two dual machine translation tasks \nop{(e.g., English -> French and French -> English)}{(e.g., En -> Fr and Fr -> En)}. However, none of the existing frameworks are directly applicable to accomplish our goal (i.e., to train a code annotation model \nop{so that it can generate annotations that helps}for generating annotations that can be utilized for code retrieval).} %For example, our code annotation model is designed to assist the code retrieval model, so this is no a minimax game setting in GANs. On the other hand, our two tasks are not dual, so the Dual Learning framework cannot help. %These two models are jointly optimized in a minimax grame formulation. 

{Figure~\ref{fig:actor_critic_framework} shows our reinforcement learning-based CoaCor framework. In the training phase, we first {train a CR model based on \nop{on} <natural language query, code snippet> {(QC)} pairs (referred to as QC-based CR model)}. Then given a code snippet, the CA model generates a sequence of NL tokens as its annotation and receives a reward from the trained CR model, which measures how effectively the generated annotation can distinguish the code snippet from others\nop{ (based on the rank of the code snippet in a candidate set when using the generated annotation as a query and the trained CR model as a search tool)}.
We formulate the annotation generation process as a Markov Decision Process \cite{bellman1957markovian} and train the CA model to maximize the received reward via an advanced reinforcement learning \cite{sutton1998introduction} framework called \textit{Advantage Actor-Critic} or \textit{A2C} \cite{mnih2016asynchronous}.\nop{ As a proof-of-concept illustration to show the effectiveness of our framework, we choose a vanilla Siamese model structure \cite{bromley1994signature} for CR, and an attention-based sequence-to-sequence model structure \cite{sutskever2014sequence} for CA.} Once the CA model is trained, we use it to generate an NL annotation $N$ for each code snippet $C$ in the code base. Therefore, for each QC pair we originally have, we can derive a QN pair. We utilize the generated annotation as a second view of the code snippet to match with a query and train {another CR model based on the derived QN pairs (referred to as QN-based CR model)}. In the testing phase, given an NL query, we rank code snippets by combining their scores from \nop{QC-based and QN-based}{both the QC-based as well as QN-based} CR models, which \nop{utilizes}{utilize} both the programming content as well as the NL annotation of a code snippet.}
%match a query with

\nop{Once the CA model is trained, we use it to generate an NL annotation for each code snippet in the code base. Thus, for each <natural language query, code snippet> pair we originally have, we can derive a <natural language query, code snippet annotation> (QN) pair, where the generated code annotation serves as a second view of the code snippet to match with a query. We utilize these generated QN pairs to train another CR model, which has a similar structure as the QC-based CR model. In the testing phase, given an NL query, we rank code snippets by combining their scores from both QC-based and QN-based CR models, which consider matching a query with both the programming content as well as the NL annotation of a code snippet.}

On a widely used benchmark dataset \cite{iyer2016summarizing} and a recently collected large-scale dataset \cite{yao2018staqc}, we show that the automatically generated annotation can significantly improve the retrieval performance.
% \nop{is retrieved based on both its code content and the corresponding generated annotation. Specifically, the CR model trained on QC pairs takes as inputs the query and the content of a code snippet candidate and outputs a matching score. Meanwhile, the CR model trained on QN pairs is trained to score a code snippet by considering whether the query and the code annotation of a code snippet candidate matches with each other. The scores from both CR models are ensembled to form a final matching score for this code snippet $C$. On the benchmark dataset \cite{iyer2016summarizing}, we show that the automatically generated annotation substantially improve the CR performance by XXX\%.}%(Section \ref{sec:experiments}).}
{More impressively, without looking at the code content, the QN-based CR model trained on our generated code annotations obtains a retrieval performance comparable to one of the state-of-the-art QC-based CR models. It also surpasses other QN-based CR models trained using code annotations generated by existing CA models.}%\fromZ{\st{In comparison, }It also surpasses other QN-based CR models trained using code \st{annotations}annotations generated by existing CA models.}} %generated by our framework % MLE or BLEU-oriented RL

To summarize, our major contributions are as follows:
\begin{itemize}[noitemsep,topsep=1pt]
    \item First, we explored a novel perspective of generating useful code annotations \textit{for} code retrieval. Unlike existing work \cite{iyer2016summarizing,xing2018DeepCom,wan2018improving}, we do not emphasize the {n-gram} overlap between the generated annotation and the human-provided one\nop{{Unlike BLEU score which emphasizes lexical n-gram matching, our framework emphasizes and generates semantically aligned annotations {as shown in Figure \ref{fig:actor_critic_framework}}.}}. Instead, we examined the real usefulness of the generated annotations and developed a machine-machine collaboration paradigm{,} where a code annotation model is trained to generate annotations that can \nop{help}be used for code retrieval. \nop{Specifically, we explored a machine-machine collaboration paradigm where a code annotation model generates annotations for code snippets \add{ to be used together with code content for} improving the CR performance.}
    \item Second, in order to accomplish our goal, we developed an effective {RL-based framework with a novel rewarding mechanism, in which a code retrieval model is directly used to formulate rewards and guide the annotation generation.} \nop{To train the code annotation model towards generating NL annotations that are helpful for code retrieval, we propose a rewarding mechanism based on the code retrieval performance, instead of maximizing the likelihood or BLEU score w.r.t. the human annotation \cite{iyer2016summarizing,xing2018DeepCom,wan2018improving}.}
    %An novel and effective learning framework to train two base models. In our framework, we propose an approach to train the model using Pseudo..
    \item Last, {we conducted extensive experiments by comparing our framework with various baselines including state-of-the-art models and variants of our framework. We showed significant \nop{XX\% }improvements of code retrieval performance on both a widely used benchmark dataset and a recently collected large-scale dataset}. %We also demonstrated that code annotations generated by our framework are superior to those generated by MLE-based and BLEU-oriented RL-based code annotation models.
\end{itemize}

The rest of this paper is organized as follows. Section \ref{sec:preliminaries} introduces the background on code annotation and code retrieval tasks\nop{as well as our intuition to associate them}. Section \ref{sec:FrameworkOverview} gives an overview of our proposed framework, with algorithm details followed in Section \ref{sec:CoaCor}. Experiments are shown in Section \ref{sec:experiments}. Finally, we discuss related work and conclude in Section \ref{sec:discussions} and \ref{sec:relatedwork}.

%% file: preliminaries.tex
%\section{Preliminaries} 

%\section{Task Definition}
%\section{Preliminaries}
\section{{Background}}
\label{sec:preliminaries} \nop{Task Definition?} \nop{I feel people might think it is our task in this paper if we use that term.}
%\fromZ{is S query or annotation? should we use Q for query and S for annotation/annotation?} \fromJ{Actually, I was thinking about the same, the annotation/annotation becomes Query for CR to get reward during RL. I used S for both, I was thinking what would be a better way.}

We adopt the same definitions for code retrieval and code annotation as previous work \cite{iyer2016summarizing, chen2018neural}. Given a natural language query $Q$ and a set of code snippet candidates {$\mathds{C}$}\nop{$C_s$}, \textit{code retrieval} is to retrieve code snippets {$C^*$} $\in$ $\mathds{C}$ \nop{$C_s$} that can match with the query. On the other hand, given a code snippet \textbf{$C$}, \textit{code annotation} is to generate a natural language (NL) annotation\nop{ (i.e., annotation)} \textbf{$N^*$} which describes the code snippet appropriately. In this work, we use \textit{code search} and \textit{code retrieval} interchangeably (and same for \textit{code annotation/summary/description}).

{Formally, for a training corpus with <natural language query, code snippet> pairs, {e.g., those collected from Stack Overflow \cite{StackOverflow} by \cite{iyer2016summarizing,yao2018staqc}}, we define the two tasks as:}
%let $N_C$ be the set of all code snippets and $N_S$ be the set of all natural language annotations. $c_j$ $\in$ $N_C$, $s_j \in N_S$
~\\

\noindent \textbf{Code Retrieval (CR):} {Given an NL Query $Q$}, a {model} \nop{function} $F_r$ will be learnt to retrieve the highest scoring code snippet $C^*$ $\in$ $\mathds{C}$.\nop{$C_Q$ (where $C_Q$ is the set of code snippet candidates for $Q$ \add{and $C_Q$ may or may not be all $C_j's$}).\fromZ{can we just say $\mathds{C}$? I feel $C_Q$ is confusing and we can say it in Experiments.}}

\begin{equation}
    C^* = \argmax_{C\in \mathds{C}} F_r(Q, C)
\end{equation}

\noindent \textbf{Code Annotation (CA):} For a given code snippet $C$, the goal is to generate an NL annotation $N^*$ that maximizes a scoring function $F_a$:
%$\in$ ($N_C \ast N_S$ $\longrightarrow$ $\mathbbm{R}$)

\begin{equation}
    N^* = \argmax_N F_a(C, N)
\end{equation}

%$F_r$ and $F_a$ can be the same scoring model
\noindent {Note that {one can use the same scoring model for $F_r$ and $F_a$} as in \cite{iyer2016summarizing,chen2018neural}, but for most of the prior work \cite{gu2018deep,xing2018DeepCom,ijcai2018-314,wan2018improving}, which consider either code retrieval or {code} annotation, researchers usually develop their own models and objective functions for $F_r$ or $F_a$. In our work, we choose two {vanilla}\nop{existing} models as our base models for CR and CA, but explore a novel perspective of how to train $F_a$ so that it can generate NL annotations that \nop{help \nop{code retrieval}\fromJ{CR}}{can be used for code retrieval}. {This perspective is inspired by various machine-machine collaboration mechanisms \cite{he2016dual,wang2017irgan, tang2017question, li2018visual} where one machine learning task can help improve another.}} %stems from the motivation to 

%\add{Some caveat: if we emphasize too much about "machine-machine" collaboration, will people challenge about comparing with work on Dual Learning??}

%\fromJJ{Please mention in comments what else should be added in this portion. I am kind of out of ideas, what else can be added as general task description.}

%\subsection{Intuition for CoaCor}
\section{Framework Overview}\label{sec:FrameworkOverview}
In this section, we first introduce our intuition and give an overview of the entire framework, before diving into more details.
\subsection{Intuition behind CoaCor}
To the best of our knowledge, previous code annotation work like \cite{iyer2016summarizing, xing2018DeepCom, ijcai2018-314, wan2018improving} focused on getting a large {n-gram} overlap {between generated and human-provided} annotations. However, it is still uncertain {(and {non-trivial}\nop{hard} to test)} how helpful the generated annotations can be. Driven by this {observation}\nop{situation}, we are the first to examine the real usefulness of the generated code annotations and how they can help a relevant task, of which we choose code retrieval as an example.

Intuitively, CoaCor can be interpreted as a \textit{collaboration mechanism} between code annotation and code retrieval. {On the one hand}, the annotation produced by the CA model provides a second view of a code snippet (in addition to its programming content) to assist code retrieval. On the other hand, when the CA model is trained to \nop{help}be useful for the retrieval task, {we expect it}\nop{it can be stimulated} to produce richer and more detailed annotations, which we {verify} in experiments later. %\add{The main challenge thus lies in how to train the CA model effectively.} From the view of code retrieval,

%% file: frameworkoverview.tex
\subsection{{Overview}}

%\fromZ{To be revised...better to combine with section 4?}
%In this paper, we propose a novel actor-critic framework named CoaCor, which explores the possibility of using code annotation to improve code retrieval performance. It aims to leverage the advantages of a generative model to boost the performance of a discriminative model for Code retrieval. The key idea for the framework is described in Figure-\ref{fig:actor_critic_framework}. 
{The main challenge to realize the above intuition lies in how to train the CA model effectively.} Our key idea to address the challenge is shown in Figure~\ref{fig:actor_critic_framework}. 

We first train a {base} CR model on $<$natural language query, code snippet$>$ (QC) pairs. Intuitively, a QC-based CR model ranks a code snippet $C$ by measuring how well it matches with the given query $Q$ (in comparison with other code snippets in the code base). From another point of view, {a well-trained} QC-based CR model can work as a measurement on whether the query $Q$ describes {the code snippet $C$}\nop{the code solution $C^*$} precisely or not. Drawing inspiration from {this view}, \nop{we propose to determine whether an annotation describes its code snippet {precisely}\nop{correctly} or not using the QC-based CR model.}{we propose using the trained QC-based CR model to determine whether an annotation describes its code snippet precisely or not} and {thereby,} train the CA model to generate rich annotations to \nop{which will help }maximize \nop{a ranking}the retrieval-based reward from the CR model. Specifically, {given a code snippet $C$, the CA model generates a sequence of NL words as its annotation $N$. At the end of the sequence,} we let the trained QC-based CR model use $N$ to search for {relevant code snippets} from the code base. If $C$ can be ranked at top places, the annotation $N$ is treated as well-written and gets a high reward; otherwise, a low reward will be returned.  We formulate this generation process as the Markov Decision Process \cite{bellman1957markovian} and train the CA model with reinforcement learning \cite{sutton1998introduction} (specifically, the \textit{Advantage Actor-Critic} algorithm \cite{mnih2016asynchronous}) to maximize the retrieval-based rewards it can receive from the QC-based CR model. We elaborate {the CR and CA model} details as well as the RL algorithm for training the CA model in Section \ref{CoaCor-CR} $\sim$ \ref{sec:CA_RL}. %In our CoaCor framework, given a code snippet $C$, the CA model generates its annotation $S$ as a sequence of NL words \add{and receives }.given a code snippet $C$ and its annotation $S$ generated by the CA model,

Once the CA model is trained, in the testing phase, it generates an NL annotation $N$ for each code snippet $C$ in the code base. Now {for each $<$NL query, code snippet$>$ pair originally in the datasets, we derive an $<$NL query, code annotation$>$ (QN) pair and train another CR model based on such QN pairs.} {This QN-based CR model complements the QC-based CR model, as they respectively use the annotation and programming content of a code snippet to match with the query.}\nop{This QN-based CR model scores a code snippet based on how well its annotation matches with the query, which complements the QC-based CR model, \add{as the latter measures} how well the programming content of the code snippet matches with the query.}
% \nop{in addition to the QC-based CR model\nop{that ranks a code snippet by considering the query $Q$ and the code snippet $C$}, we learn another CR model with a corpus of $<$natural language query, code annotation $>$ pairs derived from the original QC pairs, which retrieves relevant code snippets by measuring semantic similarity between the query and code annotations. The final matching score between the query and the code snippet is a combination of the matching scores from both the QC-based and the QS-based CR model, as we will show in Section \ref{sec:ensemble}}
We finally combine the matching scores from the two CR models to rank code snippets for a given query.%In this way, the generated annotation serves as an extra view of the code snippet to match with a query

{\textit{Note that we aim to outline a general paradigm to explore the perspective of code annotation for code retrieval, where the specific model structures for the two CR models and the CA model can be instantiated in various ways. In this work, we choose one of the simplest and most fundamental deep structures for each of them. Using more complicated model structures will make the training more challenging and we leave it as future work.}}

%which will show in Section \ref{sec:ensemble}

%Specifically, CoaCor Framework has two major components: a code retrieval model and a code annotation model. The code annotation model takes as input a code snippet and generates a sequence of \add{words} (annotation), which capture important information from the given code snippet. 

%The code retrieval model takes as input, a code snippet along with its generated code annotation. It jointly encodes, code and the generated annotation, which will be used to match with a user given natural language query. The generated code annotation serves as additional information to the input code snippet, helping boost the performance of  the code retrieval model. 
%To achieve our goal of utilizing code annotation as an input to the retrieval model, we design the framework as an actor-critic model. In our framework, the actor is a code annotation model, which generates a code annotation and the critic is the code retrieval model, which takes the generated annotation as input and provides feedback (reward) to the actor. The critic through the feedback (reward) directs the actor to generate annotation, which will help boost code retrieval performance.

%% file: model.tex
\section{\underline{Co}de \underline{A}nnotation for \underline{Co}de \underline{R}etrieval} \label{sec:CoaCor}
Now we introduce model and algorithm details in our framework.

\subsection{{Code Retrieval Model}} \label{CoaCor-CR}
\begin{figure}
\includegraphics[width=0.9\linewidth]{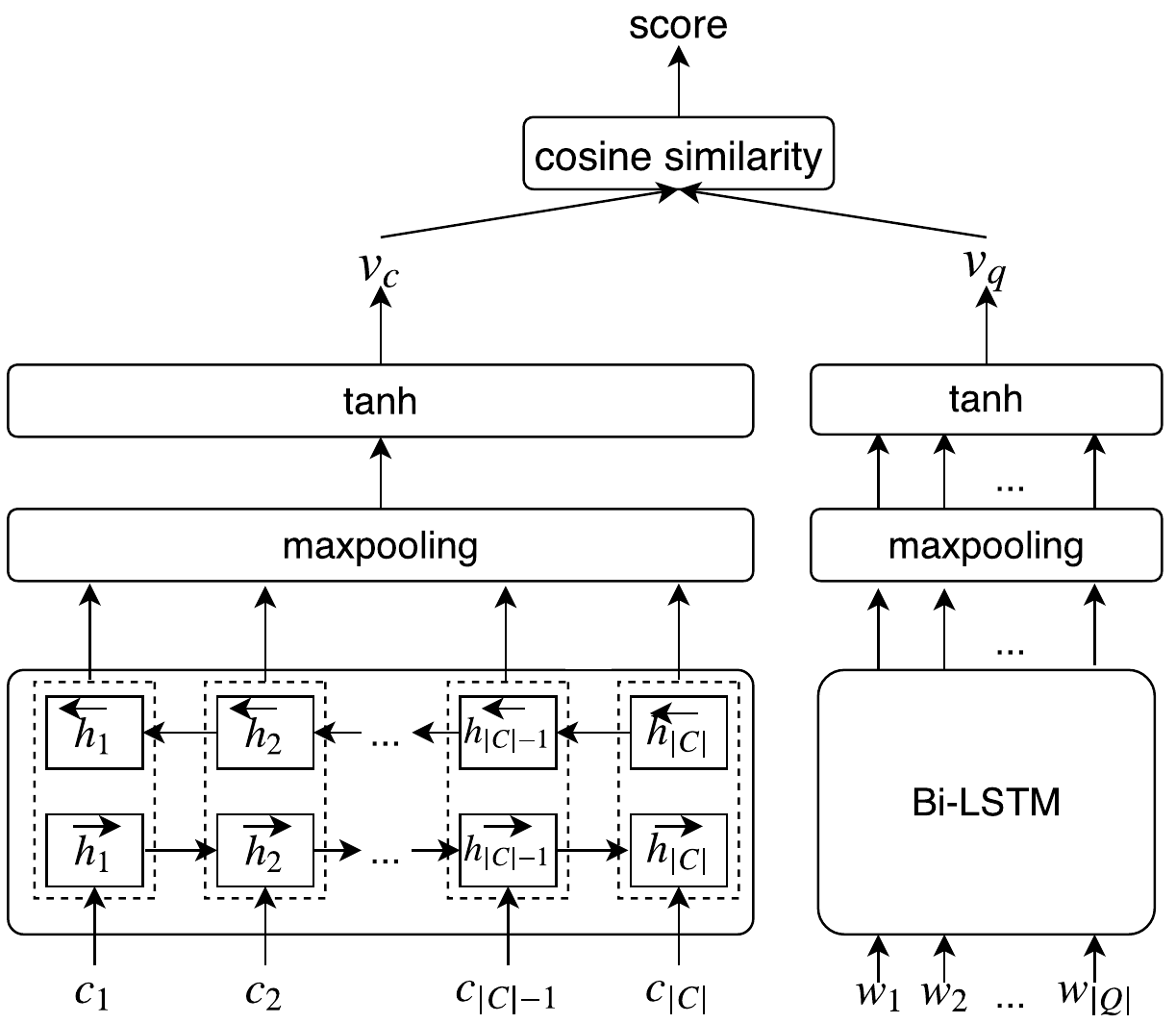}
\caption{The base code retrieval model encodes the input code snippet $\textbf{C} = (c_1, c_2, . . . , c_{|\textbf{C}|} )$  and NL {query} $\textbf{Q} = (w_1, w_2, . . . , w_{|\textbf{Q}|} )$ into a vector space and outputs a similarity score.}
\label{fig:base_code_retreival_model}
\end{figure}
% \fromH{We should keep consistent using bold or not.}\fromZ{annotation->query, add $v_q$, $v_c$, etc.}
Both QC-based and QN-based code retrieval models adopt the same deep learning structure as the previous CR work \cite{gu2018deep}. Here for simplicity, we only illustrate the QC-based CR model in detail, and the QN-based model structure is the same except that we use the generated annotation on the code snippet side.

\nop{{Generally, the QC-based CR model {takes into account the sequential nature of code snippet and NL query, and} utilizes a bidirectional Long Short-Term Memory (Bi-LSTM) network to model them{. The Bi-LSTM layer is then} followed by a max pooling layer and a tanh layer to obtain their {final} vector representations.\nop{\footnote{\label{crmodel} We also experimented with Multi-Layer Perceptron (MLP)\fromZ{we didn't try MLP, but a simple maxpooling upon word embedding layer} struture in contrast to LSTM for {encoding} similar to \cite{gu2018deep}, but it does not lead to better results.}} A similarity score is calculated using the generated code vector and the query vector, and measures how well the code snippet answers the NL query.} We elaborate the details below.}
%\fromZ{do we need this paragraph?}\fromH{we can remove it if you need space.}

As shown in Figure~\ref{fig:base_code_retreival_model}, given an NL query  $Q = w_{1 . . |Q|}$ and a code snippet $C = c_{1 . . |C|}$, we first embed the tokens of both code and NL query into vectors through a randomly initialized word embedding {matrix}, which will be learned during model training. We then use a bidirectional Long Short-Term Memory (LSTM)-based Recurrent Neural Network (RNN) \cite{hochreiter1997long,gers1999learning,medsker1999recurrent,schuster1997bidirectional,goller1996learning, zaremba2014recurrent} to learn the token representation by summarizing the contextual information from both directions. The LSTM unit is composed of three multiplicative gates.\nop{which control the proportion of information to forget and to pass on to the next time step. It is trained to selectively ``forget'' information from the hidden states, thus allowing room to take in more important information \cite{hochreiter1997long}. This property of LSTM makes it a perfect choice to encode long distance relationships in code snippets as well as NL queries.}
% \nop{LSTM has been widely used to solve semantically related tasks like speech recognition \cite{graves2013generating}, sequence tagging \cite{huang2015bidirectional,chiu2015named,ma2016end}, machine translation \cite{wu2016google,bahdanau2014neural} and etc., leading to state-of-the-art performances.}
At every time step $t$, it tracks the state of sequences by controlling how much information is updated into the new hidden state $h_t$ and memory cell $g_t$ from the previous state $h_{t-1}$, the previous memory cell $g_{t-1}$ and the current input $x_t$. On the code side, $x_t$ is the embedding vector for $c_t$, and {on the NL query side, it is the embedding vector for $w_t$}. At every time step $t$, the LSTM hidden state is updated as:
\begin{gather*}
i =  \sigma (\mathbf W_i h_{t-1} +\mathbf U_i x_t + b_i) \\ 
f = \sigma (\mathbf W_f h_{t-1} +\mathbf U_f x_t + b_f ) \\ 
o = \sigma (\mathbf W_o h_{t-1} + \mathbf U_o x_t + b_o) \\ 
g = \mathrm{tanh}(\mathbf W_g h_{t-1} +\mathbf U_g x_t + b_g) \\ 
g_t = f \odot g_{t-1} + i \odot g \\
h_t = o \odot \mathrm{tanh}(g_t)
% \tilde{c_t} = \mathrm{tanh}(\mathbf W_c h_{t-1} +\mathbf U_c x_t + b_c) \\ 
% c_t = f_t \odot c_{t-1} + i_t \odot \tilde{c_t} \\
\end{gather*}

\noindent where $\sigma$ is the element-wise sigmoid function and $\odot$ is the element-wise product.\nop{$x_t$ is the input vector (e.g. \add{word/code token embedding}) at time
$t$, and $h_t$ is the hidden state vector storing all the useful information at (and before) time $t$.}
$\mathbf U_i$, $\mathbf U_f$, $\mathbf U_g$, $\mathbf U_o$ denote the weight matrices of different gates for input $x_t$ and $\mathbf W_i$, $\mathbf W_f$, $\mathbf W_g$, $\mathbf W_o$ are the weight matrices for hidden state $h_t$, while $b_i$, $b_f$, $b_g$, $b_o$ denote the bias vectors. {For simplicity, we denote the above calculation as below (the memory cell vector $g_{t-1}$ is omitted):}\nop{Onwards, we denote the above calculation as below (the memory cell vector $g_t$ is omitted for presentation simplification):}
\begin{equation}\label{eq:lstm}
    h_t = \mathrm{LSTM}(x_t, h_{t-1})
\end{equation}

The vanilla LSTM's hidden state $h_t$ takes information from the past, knowing nothing about the future. Our CR model instead incorporates a bidirectional LSTM \cite{schuster1997bidirectional} (i.e., Bi-LSTM in Figure \ref{fig:base_code_retreival_model}), which contains a forward LSTM reading a sequence $X$ {from start to end and a backward LSTM which reads from end to start}. {The basic idea of using two directions is\nop{to present each sequence forward and backward to two separate hidden states} to capture past and future information at each step}. Then the two hidden states at each time step $t$ are concatenated to form the final hidden state $h_t$\nop{$\mathbf{h_t}$}.
\begin{gather*}
     \overset{\rightarrow}{h_t} = \mathrm{LSTM}(x_t,\overset{\rightarrow}{h_{t-1}}) \\
     \overset{\leftarrow}{h_t} = \mathrm{LSTM}(x_t,\overset{\leftarrow}{h_{t+1}}) \\
    %  \mathbf{h_t} = [\overset{\rightarrow}{h_t}, \overset{\leftarrow}{h_t}]
    h_t = [\overset{\rightarrow}{h_t}, \overset{\leftarrow}{h_t}]
\end{gather*}
{Finally, we adopt the commonly used max pooling strategy \cite{kim2014convolutional} followed by a $tanh$ layer to get the embedding vector for a sequence of length $T$.} %A typical way is to select the last hidden state $h_T$ as the embedding vector. the embedding vector of the sequence is summarized from the hidden states $h_1,..,h_T$. W
\begin{equation}
    %s = tanh(\mathrm {max pooling}([\mathbf{h_1}, ..., \mathbf{h_T}]))
    v = \mathrm{tanh}(\mathrm{maxpooling}([h_1, h_2, ..., h_T]))
\end{equation}

By applying the above encoding algorithm, we encode the code snippet $C$ and the NL query $Q$ into $v_c$ and $v_q$, respectively. Similar to \citet{gu2018deep}, to measure the relevance between \nop{the embedded vectors of }the code snippet and the query, {we use cosine similarity denoted as $cos(v_q, v_c)$.}
\nop{\begin{equation}
% cos(\vec{\pmb c}, \vec{\pmb q}) = \frac {\vec{\pmb c} \cdot \vec{\pmb q}}{||\vec{\pmb c}|| \cdot ||\vec{\pmb q}||}
cos(v_q, v_c) = \frac{(v_c)^T v_q}{||v_c|| \; ||v_q||}
\end{equation}}
\nop{where $\vec{c}$ and $\vec{q}$ are the encoding vectors of the code and the NL query respectively. }The higher the similarity, the more related the code is to the query. \nop{the vector representations are not consistent: sometimes you have the arrow and bold, but sometimes not like s above..}

{\noindent \textbf{Training.} The CR model is trained by minimizing a ranking loss similar to \cite{gu2018deep}. Specifically, for each query $Q$ in the training corpus, we prepare a triple of <$Q, C, C^-$> as a training instance, where $C$ is the correct code snippet that answers Q and $C^-$ is a negative code snippet that does not answer $Q$ (which is randomly sampled from the entire code base). The ranking loss is defined as:
\nop{At training time, we utilize ranking loss similar to \cite{gu2018deep}, and construct each training instance as a triple
<$Q, C^+, C^-$>: For each NL query $Q$, there is a positive code snippet $C^+$ (a code snippet which correctly answers $Q$) as well as a negative code snippet $C^-$ (code snippet which does not answers $Q$) \add{randomly chosen from the entire pool}. When trained on the set of <$Q, C^+, C^-$> triplets, the CR model predicts the cosine similarities of both <$Q, C^+$> and <$Q, C^-$> pairs and minimizes the ranking loss.}}

\begin{equation}\label{eq:base_cr}
    % \mathcal{L}(\theta) = \sum_{<Q, C^+, C^->'s} max(0, \epsilon - cos(q, c^+) + cos(q, c^-))
    \mathcal{L}(\theta) = \sum_{<Q, C, C^->} max(0, \epsilon - cos(v_q, v_{c}) + cos(v_q, v_{c^-}))
\end{equation}

\noindent where $\theta$ denotes the model parameters, $\epsilon$ is a constant margin, and $v_q$, $v_{c}$ and $v_{c^-}$ are the encoded vectors of $Q$, $C$ and $C^-$ respectively. Essentially, the ranking loss is a kind of hinge loss \cite{rosasco2004loss} that promotes the cosine similarity between $Q$ and $C$ to be greater than that between $Q$ and $C^-$ by at least a margin $\epsilon$. The training leads the model to project relevant queries and code snippets to be close in the vector space. 
\nop{The ranking loss encourages the cosine similarity between a NL Query and a code snippet to go up, and the cosine similarities between an NL query and incorrect code snippet to go down \cite{gu2018deep}.}

\nop{Make sure this paragraph is already paraphrased based on previous work rather than a direct copy-paste.}

% \fromH{mention this in exp: A small, fixed $\epsilon$ value of 0.05 is used in all the experiments.}
%$\mathcal{D}$ denotes the training dataset,

%\fromJJ{Where should be explain about the ranking loss and weighting technique used by different CR models?}

\subsection{{Code Annotation Model}}\label{sec:base_CA}
Formally, given a code snippet $C$, the CA model computes the probability of {generating a sequence of NL tokens {$n_{1..|N|} = (n_1, n_2, ..., n_{|N|})$} as its annotation $N$ }by:
\begin{equation}
    P(N|C) = P(n_1 | n_0, C) \prod_{t=2}^{|N|} P(n_t | n_{1 .. t-1}, C)
    \label{eqn:decodingprob}
\end{equation}
where {$n_0$} is a special token "<START>" indicating the start of the annotation generation, $n_{1..t-1} = (n_1, ..., n_{t-1})$ is the partially generated annotation till time step $t$-1, and $P(n_t | n_{1 .. t-1}, C)$ is the probability of producing $n_t$ as the next word given the code snippet $C$ and the generated $n_{1..t-1}$. {The generation stops once a special token "<EOS>" is observed.}

In this work, we choose \nop{a most}the popular sequence-to-sequence model \cite{sutskever2014sequence} as our CA model structure, {which is composed of an encoder and a decoder.} We employ the aforementioned bidirectional LSTM-based RNN structure as the encoder for code snippet $C$, and use another LSTM-based RNN {as the decoder to compute Eqn. (\ref{eqn:decodingprob})}:
\begin{gather*}
    % h_1^{\text{dec}} = \mathrm{LSTM} (n_0, h_0^{\text{dec}})\\%h_{|C|}^{\text{enc}}
    h_t^{\text{dec}} = \mathrm{LSTM} (n_{t-1}, h_{t-1}^{\text{dec}}), \forall t=1,...,|N|
\end{gather*}
where $h_t^{\text{dec}}$ is the decoder hidden state at step $t$, and $h_0^{\text{dec}}$ is initialized by concatenating the last hidden states of the code snippet encoder in both directions. In addition, a standard {global} attention layer \cite{luong2015effective} is applied in the decoder, in order to attend to important code tokens in $C$:
\begin{gather*}
\tilde{h}_t^{\text{dec}} = \mathrm{tanh} (\mathbf W_\alpha [v_{\text{attn}}, h_t^{\text{dec}}])\\
v_{\text{attn}} = \sum_{t^\prime=1}^{|C|} \alpha_{t^\prime} \; h_{t^\prime}^{\text{enc}}\\
    \alpha_{t^\prime} = \mathrm{softmax}((h_{t^\prime}^{\text{enc}})^T h_t^{\text{dec}})
\end{gather*}
where $\alpha_{t^\prime}$ is the attention weight on the $t^\prime$-th code token when generating the $t$-th word in the annotation, and $\mathbf{W}_\alpha$ is a learnable weight matrix. Finally, the $t$-th word is selected based on:
\begin{equation}\label{eq:ca_sl}
    P(n_t | n_{0..t-1}, C) = \mathrm{softmax}(\mathbf W \tilde{h}_t^{\text{dec}} + b)
\end{equation}
where $\mathbf W \in R^{|V_n| \times d}, b \in R^{|V_n|}$ project the $d$-dim hidden state $\tilde{h}_t^{\text{dec}}$ to the NL vocabulary of size $|V_n|$. 

%\subsection{Code Annotation via Reinforcement Learning} \label{sec:CA_RL}
\subsection{Training Code Annotation via RL} \label{sec:CA_RL}
\subsubsection{Code Annotation as Markov Decision Process.}
Most previous work \cite{iyer2016summarizing,xing2018DeepCom,chen2018neural} trained the CA model by maximizing the log-likelihood of generating {human annotations}, which suffers from two drawbacks:{ (1) The \textit{exposure bias} issue \cite{ranzato2015sequence,bahdanau2016actor,rennie2017self}. That is, during training, the model predicts the next word given the ground-truth annotation prefix, while at testing time, it generates the next word based on previous words generated by itself. This mismatch \nop{in word distributions }between training and testing may result in error accumulation in testing phase. (2) More importantly, maximizing the likelihood {is not aligned with our goal}\nop{cannot guarantee the optimization of the model towards} to produce annotations \nop{helpful}{that can be \nop{used}{useful}} for code retrieval.} %which is the aim of this work. during generation 

To address the above issues, we propose to formulate code annotation within the reinforcement learning (RL) framework \cite{sutton1998introduction}. Specifically, the annotation generation process is viewed as a Markov Decision Process (MDP) \cite{bellman1957markovian} consisting of four main components: 

\vspace{1mm}
\noindent \textbf{State.} At step $t$ during decoding, a state $s_t$ maintains the source code snippet and the previously generated words $n_{1..t-1}$, i.e., $s_t = \{C, n_{1..t-1}\}$. In particular, the initial decoding state $s_0 = \{C\}$ only contains the given code snippet $C$. In this work, we take the hidden state vector $\tilde{h}_{t}^{dec}$ as the vector representation of state $s_t$, and the MDP is thus processing on a continuous and infinite state space.
%\fromH{Which one is better? $s$ for state?} \fromJJ{I think s should be used for state, sometimes I myself was reading the s as state}

\vspace{1mm}
\noindent \textbf{Action.} The CA model decides the next word (or \textit{action}) $n_t \in V_n$, where $V_n$ is the NL vocabulary. Thus, the action space in our formulation is the NL vocabulary.

\vspace{1mm}
\noindent \textbf{Reward.} As introduced in Section \ref{sec:FrameworkOverview}, to encourage it to generate words \nop{helpful}{useful} for the code retrieval task, the CA model is rewarded by a {well-trained} QC-based CR model based on whether {the code snippet $C$ can be ranked at the top positions if using the generated \textit{complete} annotation $N$ as a query}\nop{the generated \textit{complete} annotation $N$ can \add{retrieve} the code snippet $C$ at the top positions or not}. Therefore, we define the reward at each step $t$ as:
\begin{equation*}
    r(s_t, n_{t}) = 
    \begin{cases}
        \text{{RetrievalReward}}(C, n_{1..t}) & \text{if } n_{t} = \text{<EOS>}\\
        0 & \text{otherwise}
    \end{cases}
\end{equation*}
where\nop{<EOS> is a special token indicating the end of the generation and $\text{RankingReward}(C, {N})$ is the ranking reward from the QC-based CR model.} {we use the popular ranking metric Mean Reciprocal Rank \cite{voorhees1999trec} (defined in Section \ref{evalmetric}) as the $\text{RetrievalReward}(C, {N})$ value.}\nop{Show we mention what a ranking reward is?} Note that, in this work, we let the CR model give a valid reward only when the annotation generation {stops, and assign a zero reward to intermediate steps. However, other designs such as giving rewards to a partial generation\nop{all generation prefixes} with the reward shaping technique \cite{bahdanau2016actor} can be reasonable and explored in the future}. %is completed
%are optional and can be explored in the future. 

\vspace{1mm}
\noindent \textbf{Policy.} The policy function $P(n_{t}|s_t)$\nop{Check the symbols here $F_a$ is used previously}\nop{inputs} {takes as input} the current state $s_t$ and outputs the probability of generating $n_{t}$ as the next word. Given our definition about state $s_t$ and {Eqn.} (\ref{eq:ca_sl}),\nop{use Eqn. not Eq}
\begin{equation}\label{eq:policy}
    P(n_{t} | s_t) = P(n_{t} | n_{1..t-1}, C) = \mathrm{softmax}(\mathbf W \tilde{h}_t^{dec} + b)
\end{equation}
Here, the policy function is stochastic in that the next word can be sampled according to the probability distribution, which allows \nop{\st{for}}action space exploration and\nop{\st{that the policy function}} can be optimized using policy gradient methods \cite{sutton1998introduction}.

The objective of the CA model training is to find a policy function $P(N|C)$ that maximizes the expected accumulative future reward:
\begin{equation} \label{eq:actor_obj}
    \max_{\phi} \mathcal{L}(\phi) = \max_{\phi} \mathds{E}_{N \sim P(\cdot | C; \phi)}[R(C, N)]
\end{equation}
where $\phi$ is the parameter of $P$ and $R(C, N) = \sum_{t=1}^{|N|} r(s_t, n_t)$ is the accumulative future reward (called ``return'').

The gradient of the above objective is derived as below.
\begin{equation} \label{eq:actor_obj_grad}
\begin{split}
\nabla_\phi \mathcal{L}(\phi) &= \mathds{E}_{N \sim P(\cdot | C;\phi)}[R(C,N) \nabla_\phi \log P(N|C; \phi)]\\
    &= \mathds{E}_{n_{1..|N|} \sim P(\cdot | C;\phi)}[\sum_{t=1}^{|N|} R_t(s_t, n_t) \nabla_\phi \log P(n_t|n_{1..t-1}, C; \phi)]
\end{split}
\end{equation}
where $R_t(s_t, n_t) = \sum_{t^\prime \geq t} r(s_{t^\prime}, n_{t^\prime})$ is the return for generating word $n_t$ given state $s_t$\nop{\st{at time step $t$}}.
%when taking action $s_t$ given the state $o_t$ at step $t$

\subsubsection{Advantage Actor-Critic for Code Annotation.}
Given the gradient in {Eqn.} (\ref{eq:actor_obj_grad}), the objective in {Eqn.} (\ref{eq:actor_obj}) can be optimized by policy gradient approaches such as REINFORCE \cite{williams1992simple} and Q-value Actor-Critic algorithm \cite{sutton2000policy, bahdanau2016actor}. However, such methods may yield very high variance when the action space (i.e., the NL vocabulary) is large and suffer from biases\nop{\nop{from} \add{against}} {when estimating the return of rarely taken actions} \cite{nguyen2017reinforcement}, leading to unstable training. To tackle the challenge, we resort to the more advanced \textit{Advantage Actor-Critic} or \textit{A2C} algorithm \cite{mnih2016asynchronous}, which has been adopted in other sequence generation tasks \cite{nguyen2017reinforcement,wan2018improving}. Specifically, the gradient function in {Eqn.} (\ref{eq:actor_obj_grad}) is replaced by an \textit{advantage} function:
\begin{gather*}
    \nabla_\phi \mathcal{L}(\phi) = \mathds{E}[\sum_{t=1}^{|N|} A_t \nabla_\phi \log P(n_t|n_{1..t-1}, C; \phi)] \stepcounter{equation}\tag{\theequation}\label{eq:a2c_obj}\\
    A_t = R_t(s_t, n_t) - V(s_t)\\
    V(s_t) = \mathds{E}_{\hat{n}_{t} \sim P(\cdot|n_{1..t-1}, C)}[R_t(s_t, \hat{n}_t)]
\end{gather*}
where $V(s_t)$ is the state value function that estimates the future reward given the current state $s_t$. Intuitively, $V(s_t)$ works as a \textit{baseline} function \cite{williams1992simple} to help the model assess its action $n_t$ more precisely: When advantage $A_t$ is greater than zero, it means that {the return for} taking action $n_t$ is better than the ``average'' return over all possible actions, given state $s_t$; otherwise, action $n_t$ performs worse than the average.

Following previous work \cite{mnih2016asynchronous,nguyen2017reinforcement, wan2018improving}, we approximate $V(s_t)$ by learning another model $V(s_t; \rho)$ parameterized by $\rho$, and the RL framework thus contains two components: the policy function $P(N|C; \phi)$ that generates the annotation (called ``Actor''), and the state value function $V(s_t; \rho)$ that approximates the return under state $s_t$ (called ``Critic''). Similar to the actor model (i.e., the{\nop{base}} CA model in Section \ref{sec:base_CA}), we train a separate attention-based sequence-to-sequence model as the critic network. The critic value is finally computed by:
\begin{equation}
    V(s_t;\rho) = \mathbf{w}^T_\rho (\tilde{h}_{t}^{dec})_{crt}+ b_\rho
\end{equation}
where $(\tilde{h}_{t}^{dec})_{crt} \in R^d$ is the critic decoder hidden state at step $t$, and $\mathbf{w}_\rho \in R^d$, $b_\rho \in R$ are trainable parameters that project the hidden state to a scalar value (i.e., the estimated state value).

The critic network is {trained} to minimize the Mean Square Error between its estimation and the true state value:
\begin{equation}\label{eq:critic}
    \min_{\rho} \mathcal{L}(\rho) = \min_{\rho} \mathds{E}_{N \sim P(\cdot|C)} [\sum_{t=1}^{|N|} (V(s_t; \rho) - R(C,N))^2]
\end{equation}
{The entire training procedure of CoaCor is shown in Algorithm \ref{alg:training}.}\nop{Should we mention what MRR is somewhere before? }

\begin{algorithm}
\begin{algorithmic}[1]
    \Require <NL query, code snippet> (QC) pairs in training set, number of iterations $E$. 
    \State Train a base code retrieval model\nop{\st{($\theta$)}} based on QC\nop{<NL query, code>} pairs, according to Eqn. (\ref{eq:base_cr}).
    \State Initialize a base code annotation model ($\phi$) and pretrain it via {MLE} according to Eqn. (\ref{eq:ca_sl}), {using $Q$ as the desired $N$ for $C$}.
    \State Pretrain a critic network ($\rho$) according to Eqn. (\ref{eq:critic}).
    \For{$iteration=1$ to $E$}
        \State Receive a code snippet $C$.
        \State Sample an annotation $N \sim P(\cdot | C; \phi)$ according to Eqn. (\ref{eq:policy}).
        \State Receive the final reward $R(C, N)$.
        \State Update the code annotation model ($\phi$) using Eqn. (\ref{eq:a2c_obj}).
        \State Update the critic network ($\rho$) using Eqn. (\ref{eq:critic}).
    \EndFor
\end{algorithmic}
    \caption{: {Training Procedure} for CoaCor\nop{-RL\textsuperscript{MRR}}. \nop{Should we also say Input: QC pairs. We do not need to specifically say output, but for input, I guess we can point out those are the only input.}\nop{is it confusing to say ``train CA on QC pairs''?}\nop{Did we define E somewhere?}}%\st{Learning algorithm}
    \label{alg:training}
\end{algorithm}

\begin{algorithm}
\begin{algorithmic}[1]
    \Require NL query $Q$, code snippet candidate $C$.
    \Ensure The matching score, $\mathrm{\textsf{score}}(Q, C)$.
    \State Receive $\mathrm{\textsf{score}}_1(Q, C) = cos(v_q, v_c)$ from a QC-based code retrieval model.
    \State Generate a code annotation $N \sim P(\cdot | C; \phi)$ via greedy search, according to Eqn. (\ref{eq:policy}). 
    \State Receive $\mathrm{\textsf{score}}_2(Q, C) = cos(v_q, v_n)$ from a QN-based code retrieval model.
    \State Calculate $\mathrm{\textsf{score}}(Q, C)$ according to Eqn. (\ref{eq:ensemble}).
\end{algorithmic}
    \caption{: {Generated Annotations for Code Retrieval.}}
    \label{alg:testing}
\end{algorithm}

% \vspace{-1cm}
%\subsection{Using Generated Annotations for Code Retrieval} \label{sec:ensemble}
\subsection{Generated Annotations for Code Retrieval} \label{sec:ensemble}
As previously shown in Figure \ref{fig:actor_critic_framework}, in the testing phase, we utilize the generated annotations to assist the code retrieval task. {Now we detail the procedure in Algorithm \ref{alg:testing}.} Specifically, for each <NL query, code snippet> pair (i.e., QC pair) in the dataset, we first derive an <NL query, code annotation> pair (i.e., QN pair) using the code annotation model. We then build another code retrieval (CR) model based on the QN pairs in our training corpus. In this work, {for simplicity}, we choose the same structure as the QC-based CR model (Section \ref{CoaCor-CR}) to {match QN pairs. However, more advanced methods for modeling the semantic similarity between two NL sentences (e.g., previous work on NL paraphase detection \cite{he2016pairwise,lan2018neural}) are applicable and can be explored as future work.}

The final matching score between query $Q$ and code snippet $C$ combines those from the QN-based and QC-based CR model:
\begin{equation} \label{eq:ensemble}
    \mathrm{score}(Q, C) = \lambda * cos(v_q, v_n) + (1 - \lambda) * cos(v_q, v_c)
\end{equation}
where $v_q, v_c, v_n$ are the encoded vectors of $Q$,  $C$, and the code annotation $N$ {respectively}. $\lambda \in [0, 1]$ is a weighting factor for the two scores {to be tuned on the validation set}.

%The detailed procedure is shown in Algorithm \ref{alg:testing}.

%% file: exp.tex
\section{Experiments} \label{sec:experiments}

In this section, we conduct extensive experiments and compare our framework with various models to show its effectiveness. 
\subsection{Experimental Setup} \label{experimentalsetup}

\noindent \textbf{Dataset.} \label{dataset}
\textbf{(1)} We experimented with the \textbf{StaQC} dataset presented by \citet{yao2018staqc}. The dataset contains 119,519 SQL <question title, code snippet> pairs mined from Stack Overflow \cite{StackOverflow}, making itself the largest-to-date in SQL domain. In our code retrieval task, the question title is considered as the NL query $Q$, which is paired with the code snippet $C$ to form the QC pair. We randomly selected 75\% of the pairs for training, 10\% for validation (containing 11,900 pairs), and the left 15\% for testing (containing 17,850 pairs). As mentioned in \cite{yao2018staqc}, the dataset may contain multiple code snippets for the same NL query. We examine the dataset split to ensure that alternative relevant code snippets for the same query would not be sampled as negative code snippets when training the CR model. {For pretraining the CA model, we consider the question title as a code annotation $N$ and form QN pairs accordingly.} \textbf{(2)} \citet{iyer2016summarizing} collected two small sets of SQL code snippets (called ``\textbf{DEV}'' and ``\textbf{EVAL}'' respectively) from Stack Overflow for validation and testing. In addition to the originally paired question title, each code snippet is manually annotated by two different NL descriptions. Therefore, in total, each set contains around 100 code snippets with three NL descriptions (resulting in around 300 QC pairs). We use them as additional datasets for model comparison.\footnote{Previous work \cite{iyer2016summarizing, chen2018neural} used only one of the three descriptions while we utilize all of them to enrich and enlarge the datasets for a more reliable evaluation.} Following \cite{iyer2016summarizing}, QC pairs occuring in DEV and EVAL set or being used as negative code snippets by them are removed from the StaQC training set.
\nop{ (2) \citet{iyer2016summarizing} provided two small sets of SQL code snippets with human provided NL annotations for validation and testing. We also used them as a second dataset for model comparison.} 

\vspace{1mm}
\noindent \textbf{Data Preprocessing.} \label{datapreprocessing}
We followed \cite{iyer2016summarizing} to perform code tokenization, which replaced table/column names with placeholder tokens and numbered them to preserve their dependencies. For text tokenization, we utilized the "word\_tokenize" tool in the NLTK toolkit \cite{bird2004nltk}. All code tokens and NL tokens with a frequency of less than 2 were replaced with an <UNK> token, resulting in totally 7726 code tokens and 7775 word tokens in the vocabulary. The average lengths of the NL query and the code snippet are 9 and 60 respectively.
%\fromH{Is it necessary to have a table with stats?}\fromZ{I think no need}}

\subsection{Evaluation} \label{evalmetric}
We evaluate a model's retrieval performance on four datasets: the validation (denoted as ``StaQC-val'') and test set (denoted as ``StaQC-test'') from StaQC \cite{yao2018staqc}, and the DEV and EVAL set from \cite{iyer2016summarizing}. For each <NL query $Q$, code snippet $C$> pair (or QC pair) in a dataset, we take $C$ as a positive code snippet and randomly sample $K$ negative code snippets from {all others except $C$ in the dataset}\nop{the code base}\nop{what do you mean by "code base"? code snippets in training set?},\footnote{{For DEV and EVAL, we use the same negative examples as \cite{iyer2016summarizing}.}} and calculate the rank of $C$ among the $K+1$ candidates. {We follow \cite{iyer2016summarizing, chen2018neural, yao2018staqc} to set $K=49$.} The retrieval performance of a model is then assessed by the Mean Reciprocal Rank (MRR) metric \cite{voorhees1999trec} over the entire set $\mathcal{D} = \{(Q_1,C_1), (Q_2,C_2), ..., (Q_{|\mathcal{D}|},C_{|\mathcal{D}|})\}$:
$$MRR = \frac{1}{|\mathcal{D}|} \sum_{i=1}^{|\mathcal{D}|} \frac{1}{Rank_i}$$
where $Rank_i$ is the rank of $C_i$ for query $Q_i$. The higher the MRR value, the better the code retrieval performance.

\subsection{Methods to Compare} \label{baselines}
In order to test the effectiveness of our CoaCor framework, we compare it with both existing baselines and our proposed variants. 

\vspace{1mm}
\noindent \textbf{Existing Baselines}. We choose the following state-of-the-art code retrieval models, which are based on QC pairs, for comparison. 

%\label{DCS}
\begin{itemize}
\item \text{Deep Code Search (DCS) \cite{gu2018deep}.} The original DCS model \cite{gu2018deep} adopts a similar structure as Figure \ref{fig:base_code_retreival_model} for CR in Java domain. To learn the vector representations for code snippets, in addition to code tokens, it also considers features like function names and API sequences, all of which are combined into a fully connected layer. In our \nop{task setting}{dataset}, we do not {have} these features, and thus slightly modify their {original} model to be the same as our QC-based CR model (Figure \ref{fig:base_code_retreival_model}). 

\item \text{CODE-NN \cite{iyer2016summarizing}.} CODE-NN is one of the state-of-the-art models for both code retrieval and code annotation. Its core component is an LSTM-based RNN with an attention mechanism, which models the probability of generating an NL sentence conditioned on a given code snippet. For code retrieval, given an NL query, CODE-NN computes the likelihood of generating the query as an annotation for each code snippet and ranks code snippets based on the likelihood.
\end{itemize}

\begin{table*}[ht!]
    \centering
    \begin{tabular}{|c|c|c|c|c|}
    \hline
        Model & DEV & EVAL & StaQC-val & StaQC-test \\\hline
        \hline
        \multicolumn{5}{|c|}{Existing (QC-based) CR Baselines} \\\hline
        DCS \cite{gu2018deep} & 0.566 & 0.555 & 0.534 & 0.529 \\
        CODE-NN \cite{iyer2016summarizing} & 0.530 & 0.514 & 0.526 & 0.522 \\
        \hline
        \hline
        \multicolumn{5}{|c|}{QN-based CR Variants} \\\hline
        QN-CodeNN & 0.369 & 0.360 & 0.336 & 0.333 \\
        QN-MLE & 0.429 & 0.411 & 0.427 & 0.424 \\
        QN-RL\textsuperscript{BLEU} & 0.426 & 0.402 & 0.386 & 0.381 \\
        QN-RL\textsuperscript{MRR} (ours)& 0.534 & 0.512 & 0.516 & 0.523 \\\hline
        \hline
        \multicolumn{5}{|c|}{Ensemble CR Variants} \\\hline
        QN-CodeNN + \nop{QC-}DCS\nop{CoaCor-CodeNN} & 0.566 & 0.555 & 0.534 & 0.529 \\
        QN-MLE + \nop{QC-}DCS\nop{CoaCor-SL} & 0.571 & 0.561 & 0.543 & 0.537 \\
        QN-RL\textsuperscript{BLEU} + \nop{QC-}DCS\nop{CoaCor-RL\textsuperscript{BLEU}} & 0.570 & 0.559 & 0.541 & 0.534 \\
        QN-RL\textsuperscript{MRR} + \nop{QC-}DCS\nop{CoaCor-RL\textsuperscript{MRR}} (ours)& {0.582$^*$} & \textbf{0.572$^*$} & {0.558$^*$} & {0.559$^*$} \\\hline
        \hline
        QN-RL\textsuperscript{MRR} + CODE-NN (ours) & \textbf{0.586$^*$} & {0.571$^*$} & \textbf{0.575$^*$} & \textbf{0.576$^*$} \\\hline
    \end{tabular}
    \caption{The main code retrieval results (MRR). * denotes significantly different from DCS \cite{gu2018deep} in one-tailed t-test (p < 0.01).}
    \label{tab:experiment_result}
    % \vspace{-3mm}
\end{table*}

\vspace{1mm}
\noindent \textbf{QN-based CR Variants}. {As discussed in Section \ref{sec:ensemble}, a trained CA model is used to annotate each code snippet $C$ in our datasets with an annotation $N$. The resulting <NL query $Q$, code annotation $N$> pairs\nop{ (i.e., QN pairs)} can be used to train a QN-based CR model. Depending on how \nop{to}{we} train the CA model, we have the following variants:}
\begin{itemize}
    \item QN-MLE. Similar to most previous work \cite{iyer2016summarizing, xing2018DeepCom,chen2018neural}, we simply train the CA model in the standard {MLE manner}, i.e., by maximizing the likelihood of a human-provided annotation.
    \item QN-RL\textsuperscript{BLEU}. {As introduced in Section \ref{sec:intro}, \citet{wan2018improving} proposed to train the CA model via reinforcement learning with BLEU scores \cite{papineni2002bleu} as rewards. We compare this variant with our rewarding mechanism.}
    \item QN-RL\textsuperscript{MRR}. {In our CoaCor framework, we propose to train the CA model using retrieval rewards from a QC-based CR model (see Section \ref{sec:FrameworkOverview}). Here we use the MRR score as the retrieval reward.}
\end{itemize}
Since CODE-NN \cite{iyer2016summarizing} can be used for code annotation as well, we also use its generated code annotations to train a QN-based CR model, denoted as ``QN-CodeNN''.\footnote{There is another recent code annotation method named DeepCom \cite{xing2018DeepCom}. We did not include it as baseline, since it achieved a similar performance as our MLE-based CA model (see Table~\ref{tab:bleu}) when evaluated with the standard BLEU script by \cite{iyer2016summarizing}.}

\vspace{1mm}
\noindent \textbf{Ensemble CR Variants}. As introduced in Section \ref{sec:ensemble}, we tried ensembling the QC-based CR model and the QN-based CR model to improve the retrieval performance. {We choose the DCS structure as the QC-based CR model as mentioned in Section \ref{CoaCor-CR}}. Since different QN-based CR models can be applied, we present the following 4 variants: (1) QN-MLE + \nop{QC-}DCS, (2) QN-RL\textsuperscript{BLEU} + \nop{QC-}DCS, (3) QN-RL\textsuperscript{MRR} + \nop{QC-}DCS, and (4) QN-CodeNN + \nop{QC-}DCS, where QN-MLE, QN-RL\textsuperscript{BLEU}, QN-RL\textsuperscript{MRR}, and QN-CodeNN have been introduced.

% \noindent \textbf{Implementation Details.}
\subsection{Implementation Details}
\nop{shorten it..For example.. and maybe you can say "code, preprocessed datasets and more implementation details will be available online (give an anonymous link)". Just cover the key things in CR and CA models.}
\nop{\st{We trained the base code retrieval model (i.e., DCS) on the <question title, code snippet> pairs of the StaQC training set, where the maximum lengths of queries and code snippets were set to 20 and 120, respectively. We use a special symbol <PAD> to pad the shorter sequences and the longer sequences will be restricted to the maximum length. }}

{Our implementation is based on Pytorch \cite{paszke2017automatic}.}
For CR models, we set the embedding size of words and code tokens to 200, and chose batch size in \{128, 256, 512\}, LSTM unit size in \{100, 200, 400\} and dropout rate \cite{srivastava2014dropout} in \{0.1, 0.35, 0.5\}. A small, fixed $\epsilon$ value of 0.05 is used in all the experiments. {Hyper-parameters} for each model were chosen based on the DEV set. For CODE-NN baseline, we followed \citet{yao2018staqc} to use the same model hyper-parameters as the original paper, except that the dropout rate is tuned in \{0.5, 0.7\}. The StaQC-val set was used to decay the learning rate and the best model parameters were decided based on the retrieval performance on the DEV set.

{For CA models, the embedding size of words and code tokens and the LSTM unit size were selected from \{256, 512\}. The dropout rate is selected from \{0.1, 0.3, 0.5\} and the batch size is 64. We updated model parameters using the Adam optimizer \cite{kingma2014adam} with learning rate 0.001 for MLE training and 0.0001 for RL training. The maximum length of the generated annotation is set to 20. For CodeNN, MLE-based and RL\textsuperscript{BLEU}-based CA models, the best model parameters were picked based on the model's BLEU score on DEV, while for RL\textsuperscript{MRR}-based CA model, we chose the best model according to its MRR reward on StaQC-val. For RL models, after pretraining the actor network via MLE, we first pretrain the critic network for 10 epochs\fromJ{,} then jointly train the two networks for 40 epochs. Finally, for ensemble variants, the ensemble weight $\lambda$ {in all variants} is selected from $0.0 \sim 1.0$ based on its performance on DEV.}

\subsection{Results} \label{results}
\nop{Table \ref{tab:experiment_result} shows the code retrieval evaluation results. }
{To understand our CoaCor framework, we first show several concrete examples to understand the differences between annotations generated by our model and by baseline/variant models, and then focus on {two} research questions ({RQs})\nop{ regarding the code retrieval performance}:}
\begin{itemize}
    %\item \add{{\textbf{RQ1}: How different do the code annotations generated by our proposed MRR-based code annotation model look ? }}
    \item \textbf{RQ1 (CR improves CA)}: Is the proposed retrieval reward-driven CA model capable of generating rich code annotations that can be used for code retrieval (i.e., can represent the code snippet and distinguish it from others)? 
    \item \textbf{RQ2 (CA improves CR)}: {Can the generated annotations {further} {improve existing} QC-based code retrieval models?}\nop{Are the generated code annotations \nop{useful}\fromZ{helpful} for existing code retrieval models?}
\end{itemize}
%Based on Table \ref{tab:experiment_result}, 

\subsubsection{Qualitative Analysis} \label{subsec:qualitative}
Table \ref{tab:examples} presents two examples of annotations generated by each CA model. Note that we do not target at human language-like annotations; rather, we focus on annotations that can describe/capture the functionality of a code snippet. In comparison {with baseline CA models}, our proposed RL\textsuperscript{MRR}-based CA model\nop{apparently} can produce more concrete and precise descriptions for corresponding code snippets. As shown in Example 1, the annotation generated by RL\textsuperscript{MRR} covers more conceptual keywords semantically aligned with the three NL queries (e.g., ``average'', ``difference'', ``group''), while the baseline CODE-NN and the variants generate short descriptions covering a very limited amount of conceptual keywords (e.g., without mentioning the concept ``subtracting'').

{We also notice that our CA model can generate different forms of a stem word (e.g., ``average'', ``avg'' in Example 1), partly because the retrieval-based reward tends to make the generated annotation semantically aligned with the code snippet and these diverse forms of words can help strengthen such semantic alignment and benefit the code retrieval task when there are various ways to express user search intent. }

\begin{table}[ht!]
    \centering
    % \begin{tabularx}{\columnwidth}{|c|c|}
    \begin{tabular}{|p{2cm}|p{6cm}|}
    \hline
        Model & Annotation \\
    \hline\hline
        \multicolumn{2}{|c|}{Example 1 from EVAL set}\\\hline
        % SQL code & {\small WITH Fruits AS (SELECT\newline
        %                 \hspace*{2mm}CASE \newline
        %                   \hspace*{4mm}WHEN m.str LIKE `\%APPLE\%' THEN `Apple'\newline
        %                   \hspace*{4mm}WHEN m.str LIKE `\%ORANGE\%' THEN `Orange'\newline
        %                 \hspace*{2mm}END AS FruitType\newline         
        %                 \hspace*{2mm}FROM MESA m\newline
        %                 \hspace*{2mm}WHERE m.str LIKE `\%FRUIT\%')\newline
        %             SELECT FruitType, COUNT(*) FROM Fruits\newline
        %             WHERE FruitType IN (`Apple', `Orange')\newline
        %             GROUP BY FruitType;}  \\\hline
        % Human-provided & {\small sql: is it possible to `{group by}' \hl{according to} `\hl{like}' function's results?}\\\hline
        % CODE-NN & {\small how do i get the count of a column in sql?}\\\hline
        % MLE & {\small sql query to count the number of words in a specific category}\\\hline
        % RL\textsuperscript{BLEU} & {\small count in sql query}\\\hline
        % RL\textsuperscript{MRR} & {\small cte {count} {counting} \hl{like case} in db2 select \hl{condition where} or clause subquery sql server multiple recursive \hl{conditional statement} using}\\
        SQL Code & {\small SELECT col3, Format(Avg([col2]-[col1]),"hh:mm:ss")\newline 
        AS TimeDiff FROM Table1\newline
        GROUP BY col3;}\\\hline
        Human-provided & {\small (1) find the \hl{average} time in hours , mins and seconds between 2 values and show them in \hl{groups} of another column\newline
        (2) group rows of a table and find \hl{average difference} between them as a \hl{formatted date}\newline
        (3) ms access \hl{average} after \hl{subtracting}}\\\hline
        CODE-NN & {\small how do i get the \hl{average} of a column in sql?}\\\hline
        % MLE & {\small how to calculate \hl{average} of a column in sql?}\\\hline
        MLE & {\small how to get \hl{average} of the \hl{average} of a column in sql}\\\hline
        % RL\textsuperscript{BLEU} & {\small how to \hl{average} in sql? }\\\hline
        RL\textsuperscript{BLEU} & {\small how to \hl{average} in sql query}\\\hline
        % RL\textsuperscript{MRR} & {\small cast \hl{averaging} \hl{subtract} in access \hl{group averages} having decimals mysql sql prepend \hl{average} using alias name type column select database}\\
        RL\textsuperscript{MRR} & {\small \hl{average} \hl{avg} calculating \hl{difference} day in access select distinct column value sql \hl{group} by month mysql \hl{format date} function?}\\
    \hline \hline
        \multicolumn{2}{|c|}{Example 2 from StaQC-test set}\\\hline
        % SQL code & {\small SELECT id, date, site, url FROM links\newline
        %         WHERE publish = "yes"  AND
        %         date = (\newline
        %             \hspace*{2mm} SELECT date FROM links\newline
        %             \hspace*{2mm} WHERE date < `2014/02/25' \newline 
        %             \hspace*{2mm} ORDER BY date DESC\newline
        %             \hspace*{2mm} LIMIT 1 )
        %         AND category!= `Adult'\newline
        %         ORDER BY date DESC, clicks DESC\newline
        %         LIMIT 200}   \\\hline
        % Human-provided & {\small (a) fetch past records \hl{closest} to given \hl{date} and \hl{sort}\newline
        %         (b) \hl{select 200} \hl{most} popular non-adult links with \hl{date} \hl{earlier than} 2014/02/25\newline
        %         (c) mysql and php : select all fields from the same \hl{date}}\\\hline
        % CODE-NN & {\small how do i get the \hl{maximum} value of a column?}  \\\hline
        % MLE & {\small mysql query to get the \hl{last} \hl{date} of a specific day} \\\hline
        % RL\textsuperscript{BLEU} & {\small how to \hl{date} in sql query} \\\hline
        % RL\textsuperscript{MRR} & {\small \hl{order date compare} mysql select where \hl{dates} access \hl{limit} the \hl{desc} in \hl{timestamp} or having \hl{max} \hl{datetime} column using current }\\
        SQL Code & {\small SELECT Group\_concat(DISTINCT( p.products\_id )) AS comma\_separated,\newline
                        COUNT(DISTINCT p.products\_id) AS product\_count\newline
                        FROM ... }\\\hline
        Human-provided & {\small how to \hl{count} how many \hl{comma} separated values in a \hl{group\_concat}}\\\hline
        CODE-NN & {\small how do i get the \hl{count} of distinct rows?}\\\hline
        % MLE & {\small mysql query to \hl{count} distinct values}\\\hline
        MLE & {\small mysql query to get \hl{count} of distinct values in a column}\\\hline
        % RL\textsuperscript{BLEU} & {\small how to \hl{count} in mysql query}\\\hline
        RL\textsuperscript{BLEU} & {\small how to \hl{count} in mysql sql query}\\\hline
        % RL\textsuperscript{MRR} & {\small \hl{group\_concat} distinct \hl{count} unique \hl{counting} \hl{comma} in mysql alias column access concatenate columns sql server db2 single group id ms}\\
        RL\textsuperscript{MRR} & {\small \hl{group\_concat} \hl{count} concatenate distinct \hl{comma} group mysql concat column in one row rows select multiple columns of same id result}\\
    \hline
    \end{tabular}
    \caption{{Two examples of code snippets and their annotations generated by different CA models}. ``Human-provided'' refers to (multiple) human-provided NL annotations or queries. Words semantically aligned between the generated and the human-provided annotations are highlighted.}
    \label{tab:examples}
    % \vspace{-5mm}
\end{table}

\subsubsection{Code Retrieval Performance Evaluation}
{Table \ref{tab:experiment_result} shows the code retrieval evaluation results, based on which we discuss \textbf{RQ1} and \textbf{RQ2} as below}:

\vspace{1mm}
\noindent \textbf{RQ1}: To examine whether or not the code annotations generated by a CA model can represent the corresponding code snippet in the code retrieval task, we analyze its corresponding QN-based CR model, which retrieves relevant code snippets by matching the NL query $Q$ with the code annotation $N$ generated by this CA model.
Across all of the four datasets, our proposed QN-RL\textsuperscript{MRR} model, which is based on a retrieval reward\nop{MRR}-driven CA model, achieves the best results and outperforms other QN-based CR models by a wide margin of around 0.1 $\sim$ 0.2 absolute MRR. More impressively, its performance is already on a par with the CODE-NN model, which is one of the state-of-the-art models for the code retrieval task, even though it understands a code snippet solely based on its annotation and without looking at the code content. This demonstrates that the code annotation generated by our proposed {framework}\nop{retrieval reward\nop{MRR}-driven CA model} can reflect the semantic meaning of each code snippet more precisely.

To further understand whether or not the retrieval-based reward can serve as a better reward metric than BLEU (in terms of stimulating a CA model to generate useful annotations), we present the BLEU score of each CA model in Table \ref{tab:bleu}.\footnote{{BLEU is evaluated with the script provided by \citet{iyer2016summarizing}: \url{https://github.com/sriniiyer/codenn/blob/master/src/utils/bleu.py}.}}
When connecting this table with Table \ref{tab:experiment_result}, we observe an inverse trend: 
{For the RL\textsuperscript{BLEU} model which is \textit{trained for} a higher BLEU score, although it can improve the MLE-based CA model by more than 2\% absolute BLEU {on three sets}, it harms the latter's ability on producing useful code annotations (as revealed by the performance of QN-RL\textsuperscript{BLEU} in Table \ref{tab:experiment_result}, which is worse than QN-MLE by around 0.04 absolute MRR on StaQC-val and StaQC-test). In contrast, our proposed RL\textsuperscript{MRR} model, despite getting the lowest BLEU score, is capable of generating annotations useful for the retrieval task.}
This is mainly because that BLEU score calculates surface form overlaps {while the retrieval-based reward measures the semantically aligned correspondences}.

These observations imply an interesting conclusion: {\textit{Compared with BLEU, \nop{the retrieval-based MRR score}a {(task-oriented)} semantic measuring reward, such as our retrieval-based MRR score, {can better stimulate}\nop{is a better training \nop{reward to stimulate a}stimulation for} the model to {produce detailed and useful generations}\nop{generate detailed annotations to describe the given code snippet}.}} This is in line with the recent discussions on whether the automatic BLEU score is an appropriate evaluation metric for generation tasks or not \cite{liu2016not, novikova2017we}. In our work, we study the potential to use {the performance of a relevant model} \nop{task-oriented machine performance as a reward} to guide the learning of the target model, which can be generalized to many other scenarios, e.g., conversation generation \cite{li2016deep}, machine translation \cite{bahdanau2016actor,ranzato2015sequence}, etc.

\begin{table}[t!]
    \centering
    \begin{tabular}{|c|c|c|c|c|}
    \hline
        Model & DEV & EVAL & StaQC-val & StaQC-test \\\hline\hline
        CODE-NN \cite{iyer2016summarizing} & 17.43 & 16.73 & 8.89 & 8.96  \\
        MLE & 18.99 & 19.87 & 10.52 & 10.55 \\
        RL\textsuperscript{BLEU} & 21.12 & 18.52 & 12.72 & 12.78 \\
        RL\textsuperscript{MRR} & 8.09 & 8.52 & 5.56 & 5.60 \\
    \hline
    \end{tabular}
    \caption{The BLEU score of each code annotation model.}
    \label{tab:bleu}
\end{table}

\vspace{1mm}
\noindent \textbf{RQ2}: 
We first inspect whether the generated code annotations can assist the base code retrieval model (i.e., DCS) or not by comparing \nop{the}several ensemble CR variants.
It is shown that, by simply combining the matching scores from QN-RL\textsuperscript{MRR} and DCS\nop{QC-based and QN-based CR models} with a weighting factor, our proposed\nop{\st{CoaCor}} model is able to {significantly} outperform the DCS model by 0.01 $\sim$ 0.03 and the CODE-NN baseline by 0.03 $\sim$ 0.06 consistently \nop{on}{across} all datasets, showing the advantage of utilizing code annotations for code retrieval. {Particularly}\nop{Interestingly}, the best performance is achieved {when the}\nop{for} ensemble weight $\lambda = 0.4$ (i.e., 0.4 weight on the QN-based CR score and 0.6 on the QC-based CR score), meaning that the model relies \nop{much}{heavily} on the code annotation \nop{for a}{to achieve} better performance.

{In contrast, QN-CodeNN, QN-MLE and QN-RL\textsuperscript{BLEU} can hardly improve the base DCS model, and their best performances are all achieved when the ensemble weight $\lambda = 0.0$ $\sim$ $0.2$, indicating little help from annotations generated by CODE-NN, MLE-based and BLEU-rewarded CA.}
This is consistent with our conclusions to RQ1.

We also investigate the benefit of our generated annotations to other code retrieval models (besides {DCS})\nop{. To answer this question, we experimented with} by examining a baseline ``QN-RL\textsuperscript{MRR} + CODE-NN'', which combines QN-RL\textsuperscript{MRR} and \nop{\st{QC-based}} CODE-NN (as a QC-based CR model) to score a code snippet candidate. {As mentioned in Section \ref{baselines}, CODE-NN scores a code snippet \nop{being}{by} the likelihood of generating the given NL query when taking this code snippet as the input. Since the score is in a different range from the cosine similarity given by QN-RL\textsuperscript{MRR}, we first rescale it by taking its log value and dividing it by the largest absolute log score among all code candidates. The rescaled score is then combined with the cosine similarity score from QN-RL\textsuperscript{MRR} following Eqn. (\ref{eq:ensemble}).}\nop{Maybe a bit more explanation about how you combine here? no joint training..}
The result is shown in {the last row} of Table \ref{tab:experiment_result}. It is very impressive that, {with the help of QN-RL\textsuperscript{MRR}\nop{utilizing the code annotations generated by our RL\textsuperscript{MRR} framework}}, the CODE-NN model can be improved by $\geq$ 0.05 absolute MRR value across all test sets. \nop{This result again demonstrates the advantage of the MRR rewarding mechanism.}
\nop{The result points out a promising research direction to involve \textit{multiple} models to \textit{collaborate} towards a better performance, which we will explore in the future.}

\vspace{1mm}
\noindent {\textit{In summary}, through extensive experiments, we show that our proposed framework can generate code annotations that are much more useful for building effective code retrieval models, in comparison with existing CA models or those trained by MLE or BLEU-based RL. {Additionally, the generated code annotations} can further improve the retrieval performance, when combined with existing CR models like DCS and CODE-NN.}

\section{Discussion} \label{sec:discussions}
{In this work, we propose a novel perspective of using a relevant {downstream} task (i.e., code retrieval) to guide the learning of a target task (i.e., code annotation), \nop{which illustrates} {illustrating} a novel machine-machine collaboration paradigm. It is shown that the annotations generated by the RL\textsuperscript{MRR} CA model ({trained with rewards from the DCS model}) can boost the performance of the CODE-NN model, which was not involved in \nop{the training phase at all}{any stage of the training process}.\nop{This observation inspires us} {It is interesting to} explore more about machine-machine collaboration mechanisms, where multiple models for either the same task or relevant tasks can be utilized {in tandem} to provide different views or effective rewards to improve the final performance.}

{In terms of training, we also {experimented with} directly using a QN-based {CR model} or an ensemble CR model for rewarding the CA model\nop{to generate annotation $N$ close to the query $Q$ or to directly maximize the ensemble retrieval performance}. However, these approaches do not work well, since we do not have\nop{proper QN pairs as the training data} {a rich set of QN pairs as training data in the beginning}.\nop{How to} Collecting paraphrases of queries to form QN pairs is non-trivial, which we leave to the future.} 

{Finally, our CoaCor framework is applicable to other programming languages, such as Python and C\#, extension to which is interesting to study as future work.}

%% file: relatedwork.tex
\section{Related Work} \label{sec:relatedwork}

\noindent \textbf{Code Retrieval.} 
% \nop{As introduced in Section \ref{sec:intro}, code retrieval has been explored by approaches ranging from conventional information retrieval methods \cite{haiduc2013automatic, lu2015query, hill2014nl,keivanloo2014spotting, Vinayakarao:2017:AIS:3018661.3018691} to recent deep learning models \cite{allamanis2015bimodal, iyer2016summarizing, gu2018deep}.}
As introduced in Section~\ref{sec:intro}, code retrieval has been studied widely with information retrieval methods \cite{haiduc2013automatic, lu2015query, hill2014nl,keivanloo2014spotting, Vinayakarao:2017:AIS:3018661.3018691} and recent deep learning models \cite{allamanis2015bimodal, iyer2016summarizing, gu2018deep}. Particularly, \citet{keivanloo2014spotting} extracted abstract programming patterns and their associated NL keywords from code snippets in {a} code base, with which a given NL query can be projected to a set of associated programming patterns {facilitating} code content-based search. Similarly, \citet{Vinayakarao:2017:AIS:3018661.3018691} built an entity discovery system to mine NL phrases and their associated syntactic patterns, based on which they annotated each line of code snippets with NL phrases. Such annotations were utilized to improve NL keyword-based search engines. Different from these work, we construct\nop{ed} a neural network-based code annotation model to {describe the functionality of an entire code snippet}\nop{complete code snippet with a high-level NL description}. Our code annotation model is explicitly trained to produce meaningful words that can be used for code search. In our framework, the code retrieval model adopts a similar deep structure as the Deep Code Search model proposed by \citet{gu2018deep}, which projects a NL query and a code snippet into a vector space and measures the cosine similarity between them.

\vspace{1mm}
\noindent \textbf{Code Annotation.}
{Code annotation/summarization has drawn a lot of attention in recent years. Earlier works tackled the problem using template-based approaches \cite{sridhara2010towards, moreno2013automatic} and topic n-grams models \cite{movshovitz2013natural} while the recent\nop{state-of-the-art} techniques \cite{allamanis2016convolutional, xing2018DeepCom, iyer2016summarizing, ijcai2018-314, jiang2017automatically, loyola2017neural} are mostly built upon deep neural networks.}
\nop{Code annotation/summarization has been tackled by approaches ranging from template-based \cite{sridhara2010towards, moreno2013automatic} to advanced deep neural networks \cite{allamanis2016convolutional, xing2018DeepCom, iyer2016summarizing, ijcai2018-314, jiang2017automatically, loyola2017neural}.} Specifically, \citet{sridhara2010towards} developed a software word usage model to identify action, theme and other arguments from a given code snippet, and generated code comments with templates. \citet{allamanis2016convolutional} employed convolution on the input tokens to detect local time-invariant and long-range topical attention features to summarize a code snippet into a short, descriptive function name-like {summary}. Most related to our work are\nop{\cite{iyer2016summarizing} and} \cite{xing2018DeepCom} and \cite{wan2018improving}, which utilized sequence-to-sequence networks with attention over code tokens to generate natural language annotations. They aimed to generate NL annotations as close as possible to human-provided annotations for human readability, and hence adopted the Maximum Likelihood Estimation (MLE) or {BLEU score optimization} as the objective. However, our goal is to generate code annotations \textit{which can be used for code retrieval}, and therefore we design a retrieval-based reward to drive our training.
%{I like retrieval-based reward better than "ranking rewards". What do you think?}\fromZ{yes}
%The base code annotation model in our CoaCor framework is very similar to \cite{iyer2016summarizing} and \cite{xing2018DeepCom}. 
% presented a template-based approach for comment generation based on a software word usage model

\vspace{1mm}
\noindent \textbf{Deep Reinforcement Learning for Sequence Generation.} 
{Reinforcement learning (RL) \cite{sutton1998introduction} has shown great success in various tasks where an agent has to perform multiple actions before obtaining a reward or when the metric to optimize is not differentiable. \nop{As a typical scenario,} The sequence generation tasks, such as machine translation \cite{ranzato2015sequence, nguyen2017reinforcement,bahdanau2016actor}, image captioning \cite{rennie2017self}, dialogue generation \cite{li2016deep} and text summarization \cite{paulus2017deep}, \nop{have been beneficial}{have all benefitted} from RL to address the \textit{exposure bias} issue \cite{bahdanau2016actor, rennie2017self, ranzato2015sequence} and to directly optimize the model towards a certain metric (e.g., BLEU).}
\nop{Reinforcement learning (RL) \cite{sutton1998introduction} has shown great success in sequence generation tasks, such as machine translation \cite{ranzato2015sequence, nguyen2017reinforcement,bahdanau2016actor}, image captioning \cite{rennie2017self}, dialog generation \cite{li2016deep} and text summarization \cite{paulus2017deep}.}
Particularly, \citet{ranzato2015sequence} were among the first to successfully apply the REINFORCE algorithm \cite{williams1992simple} to train RNN models for several sequence generation tasks, indicating that directly optimizing the metric used at test time can lead to significantly better models than those trained via MLE. \citet{bahdanau2016actor} additionally learned a \textit{critic} network to better estimate the return (i.e., future rewards) of taking a certain action under a specific state, and trained the entire generation model via the \textit{Actor-Critic} algorithm \cite{sutton2000policy}. We follow \citet{nguyen2017reinforcement} and \citet{wan2018improving} to further introduce an \textit{advantage} function and train the code annotation model via the \textit{Advantage Actor-Critic} algorithm \cite{mnih2016asynchronous}, which is helpful for reducing biases from rarely taken actions. However, unlike their work, the reward in our framework is {based on the performance on a different yet relevant task\nop{a different machine learning model} (i.e., code retrieval)}, rather than the BLEU metric.

\vspace{1mm}
\noindent \textbf{Machine-Machine Collaboration via Adversarial Training\\ and Dual/Joint Learning.}
Various kinds of machine-machine collaboration mechanisms have been studied in many scenarios \cite{goodfellow2014generative, he2016dual, wang2017irgan, tang2017question,li2018visual}. For example, \citet{goodfellow2014generative} proposed the Generative Adversarial Nets (GANs) framework, where a generative model generates \nop{\st{challenging}} images to fool a discriminative classifier, and the latter is further improved to distinguish the {generated from the real}\nop{noisy images from correct} ones. \citet{he2016dual} proposed the dual learning framework and jointly optimized the machine translation from English to French and from French to English.{\nop{By the time we publish this work, We also notice a recent work} \citet{D18-1421} trained a paraphrase generator by rewards from a paraphrase evaluator model.} In the context of code retrieval and annotation, \citet{chen2018neural} and \citet{iyer2016summarizing} showed that their models can be used directly or with slight modification for both tasks, but their training objective only considered one of the two tasks. All these frameworks are not directly applicable to achieve our goal, i.e., training a code annotation model to generate rich NL annotations that can be used for code search.

%% file: conclusion.tex
\section{Conclusion}

This paper explored a novel perspective of generating code annotations \textit{for} code retrieval. To this end, we proposed a reinforcement learning-based framework (named ``CoaCor'') to maximize a retrieval-based reward. Through comprehensive experiments, we demonstrated that the annotation generated by our framework is more detailed to represent the semantic meaning of a code snippet. Such annotations can also improve the existing code content-based retrieval models significantly. In the future, we will explore other usages of the generated code annotations, as well as generalizing our framework to other tasks such as machine translation.

\section*{Acknowledgments}
This research was sponsored in part by the Army Research Office under cooperative agreements W911NF-17-1-0412, NSF Grant IIS1815674, Fujitsu gift grant, and Ohio Supercomputer
Center \cite{OhioSupercomputerCenter1987}. The views and conclusions contained herein are those of the authors and should not be interpreted as representing the official policies, either expressed or implied, of the Army Research Office or the U.S. Government. The U.S. Government is authorized to reproduce and distribute reprints for Government purposes notwithstanding any copyright notice herein.

%% file: main.bbl

\begin{thebibliography}{64}

%%% the Web address.

\ifx \showCODEN    \undefined \def \showCODEN     #1{\unskip}     \fi
\ifx \showDOI      \undefined \def \showDOI       #1{#1}\fi
\ifx \showISBNx    \undefined \def \showISBNx     #1{\unskip}     \fi
\ifx \showISBNxiii \undefined \def \showISBNxiii  #1{\unskip}     \fi
\ifx \showISSN     \undefined \def \showISSN      #1{\unskip}     \fi
\ifx \showLCCN     \undefined \def \showLCCN      #1{\unskip}     \fi
\ifx \shownote     \undefined \def \shownote      #1{#1}          \fi
\ifx \showarticletitle \undefined \def \showarticletitle #1{#1}   \fi
\ifx \showURL      \undefined \def \showURL       {\relax}        \fi
% invisible to TeX
\providecommand\bibfield[2]{#2}
\providecommand\bibinfo[2]{#2}
\providecommand\natexlab[1]{#1}
\providecommand\showeprint[2][]{arXiv:#2}

\bibitem[\protect\citeauthoryear{Allamanis, Barr, Devanbu, and
  Sutton}{Allamanis et~al\mbox{.}}{2018}]%
        {allamanis2018survey}
\bibfield{author}{\bibinfo{person}{Miltiadis Allamanis},
  \bibinfo{person}{Earl~T Barr}, \bibinfo{person}{Premkumar Devanbu}, {and}
  \bibinfo{person}{Charles Sutton}.} \bibinfo{year}{2018}\natexlab{}.
\newblock \showarticletitle{A survey of machine learning for big code and
  naturalness}.
\newblock \bibinfo{journal}{\emph{ACM Computing Surveys (CSUR)}}
  \bibinfo{volume}{51}, \bibinfo{number}{4} (\bibinfo{year}{2018}),
  \bibinfo{pages}{81}.
\newblock


\bibitem[\protect\citeauthoryear{Allamanis, Peng, and Sutton}{Allamanis
  et~al\mbox{.}}{2016}]%
        {allamanis2016convolutional}
\bibfield{author}{\bibinfo{person}{Miltiadis Allamanis}, \bibinfo{person}{Hao
  Peng}, {and} \bibinfo{person}{Charles Sutton}.}
  \bibinfo{year}{2016}\natexlab{}.
\newblock \showarticletitle{A convolutional attention network for extreme
  summarization of source code}. In \bibinfo{booktitle}{\emph{International
  Conference on Machine Learning}}. \bibinfo{pages}{2091--2100}.
\newblock


\bibitem[\protect\citeauthoryear{Allamanis, Tarlow, Gordon, and Wei}{Allamanis
  et~al\mbox{.}}{2015}]%
        {allamanis2015bimodal}
\bibfield{author}{\bibinfo{person}{Miltos Allamanis}, \bibinfo{person}{Daniel
  Tarlow}, \bibinfo{person}{Andrew Gordon}, {and} \bibinfo{person}{Yi Wei}.}
  \bibinfo{year}{2015}\natexlab{}.
\newblock \showarticletitle{Bimodal modelling of source code and natural
  language}. In \bibinfo{booktitle}{\emph{International Conference on Machine
  Learning}}. \bibinfo{pages}{2123--2132}.
\newblock


\bibitem[\protect\citeauthoryear{Bahdanau, Brakel, Xu, Goyal, Lowe, Pineau,
  Courville, and Bengio}{Bahdanau et~al\mbox{.}}{2016}]%
        {bahdanau2016actor}
\bibfield{author}{\bibinfo{person}{Dzmitry Bahdanau}, \bibinfo{person}{Philemon
  Brakel}, \bibinfo{person}{Kelvin Xu}, \bibinfo{person}{Anirudh Goyal},
  \bibinfo{person}{Ryan Lowe}, \bibinfo{person}{Joelle Pineau},
  \bibinfo{person}{Aaron Courville}, {and} \bibinfo{person}{Yoshua Bengio}.}
  \bibinfo{year}{2016}\natexlab{}.
\newblock \showarticletitle{An actor-critic algorithm for sequence prediction}.
\newblock \bibinfo{journal}{\emph{arXiv preprint arXiv:1607.07086}}
  (\bibinfo{year}{2016}).
\newblock


\bibitem[\protect\citeauthoryear{Bellman}{Bellman}{1957}]%
        {bellman1957markovian}
\bibfield{author}{\bibinfo{person}{Richard Bellman}.}
  \bibinfo{year}{1957}\natexlab{}.
\newblock \showarticletitle{A Markovian decision process}.
\newblock \bibinfo{journal}{\emph{Journal of Mathematics and Mechanics}}
  (\bibinfo{year}{1957}), \bibinfo{pages}{679--684}.
\newblock


\bibitem[\protect\citeauthoryear{Biggerstaff, Mitbander, and
  Webster}{Biggerstaff et~al\mbox{.}}{1994}]%
        {biggerstaff1994program}
\bibfield{author}{\bibinfo{person}{Ted~J Biggerstaff},
  \bibinfo{person}{Bharat~G Mitbander}, {and} \bibinfo{person}{Dallas~E
  Webster}.} \bibinfo{year}{1994}\natexlab{}.
\newblock \showarticletitle{Program understanding and the concept assignment
  problem}.
\newblock \bibinfo{journal}{\emph{Commun. ACM}} \bibinfo{volume}{37},
  \bibinfo{number}{5} (\bibinfo{year}{1994}), \bibinfo{pages}{72--82}.
\newblock


\bibitem[\protect\citeauthoryear{Bird and Loper}{Bird and Loper}{2004}]%
        {bird2004nltk}
\bibfield{author}{\bibinfo{person}{Steven Bird} {and} \bibinfo{person}{Edward
  Loper}.} \bibinfo{year}{2004}\natexlab{}.
\newblock \showarticletitle{NLTK: the natural language toolkit}. In
  \bibinfo{booktitle}{\emph{Proceedings of the ACL 2004 on Interactive poster
  and demonstration sessions}}. Association for Computational Linguistics,
  \bibinfo{pages}{31}.
\newblock


\bibitem[\protect\citeauthoryear{Center}{Center}{1987}]%
        {OhioSupercomputerCenter1987}
\bibfield{author}{\bibinfo{person}{Ohio~Supercomputer Center}.}
  \bibinfo{year}{1987}\natexlab{}.
\newblock \bibinfo{title}{Ohio Supercomputer Center}.
\newblock \bibinfo{howpublished}{\url{http://osc.edu/ark:/19495/f5s1ph73}}.
\newblock


\bibitem[\protect\citeauthoryear{Chen and Zhou}{Chen and Zhou}{2018}]%
        {chen2018neural}
\bibfield{author}{\bibinfo{person}{Qingying Chen} {and}
  \bibinfo{person}{Minghui Zhou}.} \bibinfo{year}{2018}\natexlab{}.
\newblock \showarticletitle{A neural framework for retrieval and summarization
  of source code}. In \bibinfo{booktitle}{\emph{Proceedings of the 33rd
  ACM/IEEE International Conference on Automated Software Engineering}}. ACM,
  \bibinfo{pages}{826--831}.
\newblock


\bibitem[\protect\citeauthoryear{Gers, Schmidhuber, and Cummins}{Gers
  et~al\mbox{.}}{1999}]%
        {gers1999learning}
\bibfield{author}{\bibinfo{person}{Felix~A Gers}, \bibinfo{person}{J{\"u}rgen
  Schmidhuber}, {and} \bibinfo{person}{Fred Cummins}.}
  \bibinfo{year}{1999}\natexlab{}.
\newblock \showarticletitle{Learning to forget: Continual prediction with
  LSTM}.
\newblock  (\bibinfo{year}{1999}).
\newblock


\bibitem[\protect\citeauthoryear{Goller and Kuchler}{Goller and
  Kuchler}{1996}]%
        {goller1996learning}
\bibfield{author}{\bibinfo{person}{Christoph Goller} {and}
  \bibinfo{person}{Andreas Kuchler}.} \bibinfo{year}{1996}\natexlab{}.
\newblock \showarticletitle{Learning task-dependent distributed representations
  by backpropagation through structure}. In \bibinfo{booktitle}{\emph{Neural
  Networks, 1996., IEEE International Conference on}},
  Vol.~\bibinfo{volume}{1}. IEEE, \bibinfo{pages}{347--352}.
\newblock


\bibitem[\protect\citeauthoryear{Goodfellow, Pouget-Abadie, Mirza, Xu,
  Warde-Farley, Ozair, Courville, and Bengio}{Goodfellow et~al\mbox{.}}{2014}]%
        {goodfellow2014generative}
\bibfield{author}{\bibinfo{person}{Ian Goodfellow}, \bibinfo{person}{Jean
  Pouget-Abadie}, \bibinfo{person}{Mehdi Mirza}, \bibinfo{person}{Bing Xu},
  \bibinfo{person}{David Warde-Farley}, \bibinfo{person}{Sherjil Ozair},
  \bibinfo{person}{Aaron Courville}, {and} \bibinfo{person}{Yoshua Bengio}.}
  \bibinfo{year}{2014}\natexlab{}.
\newblock \showarticletitle{Generative adversarial nets}. In
  \bibinfo{booktitle}{\emph{Advances in neural information processing
  systems}}. \bibinfo{pages}{2672--2680}.
\newblock


\bibitem[\protect\citeauthoryear{Gu, Zhang, and Kim}{Gu et~al\mbox{.}}{2018}]%
        {gu2018deep}
\bibfield{author}{\bibinfo{person}{Xiaodong Gu}, \bibinfo{person}{Hongyu
  Zhang}, {and} \bibinfo{person}{Sunghun Kim}.}
  \bibinfo{year}{2018}\natexlab{}.
\newblock \showarticletitle{Deep code search}. In
  \bibinfo{booktitle}{\emph{Proceedings of the 40th International Conference on
  Software Engineering}}. ACM, \bibinfo{pages}{933--944}.
\newblock


\bibitem[\protect\citeauthoryear{Haiduc, Bavota, Marcus, Oliveto, De~Lucia, and
  Menzies}{Haiduc et~al\mbox{.}}{2013}]%
        {haiduc2013automatic}
\bibfield{author}{\bibinfo{person}{Sonia Haiduc}, \bibinfo{person}{Gabriele
  Bavota}, \bibinfo{person}{Andrian Marcus}, \bibinfo{person}{Rocco Oliveto},
  \bibinfo{person}{Andrea De~Lucia}, {and} \bibinfo{person}{Tim Menzies}.}
  \bibinfo{year}{2013}\natexlab{}.
\newblock \showarticletitle{Automatic query reformulations for text retrieval
  in software engineering}. In \bibinfo{booktitle}{\emph{Proceedings of the
  2013 International Conference on Software Engineering}}. IEEE Press,
  \bibinfo{pages}{842--851}.
\newblock


\bibitem[\protect\citeauthoryear{He, Xia, Qin, Wang, Yu, Liu, and Ma}{He
  et~al\mbox{.}}{2016}]%
        {he2016dual}
\bibfield{author}{\bibinfo{person}{Di He}, \bibinfo{person}{Yingce Xia},
  \bibinfo{person}{Tao Qin}, \bibinfo{person}{Liwei Wang},
  \bibinfo{person}{Nenghai Yu}, \bibinfo{person}{Tieyan Liu}, {and}
  \bibinfo{person}{Wei-Ying Ma}.} \bibinfo{year}{2016}\natexlab{}.
\newblock \showarticletitle{Dual learning for machine translation}. In
  \bibinfo{booktitle}{\emph{Advances in Neural Information Processing
  Systems}}. \bibinfo{pages}{820--828}.
\newblock


\bibitem[\protect\citeauthoryear{He and Lin}{He and Lin}{2016}]%
        {he2016pairwise}
\bibfield{author}{\bibinfo{person}{Hua He} {and} \bibinfo{person}{Jimmy Lin}.}
  \bibinfo{year}{2016}\natexlab{}.
\newblock \showarticletitle{Pairwise word interaction modeling with deep neural
  networks for semantic similarity measurement}. In
  \bibinfo{booktitle}{\emph{Proceedings of the 2016 Conference of the North
  American Chapter of the Association for Computational Linguistics: Human
  Language Technologies}}. \bibinfo{pages}{937--948}.
\newblock


\bibitem[\protect\citeauthoryear{Hill, Roldan-Vega, Fails, and Mallet}{Hill
  et~al\mbox{.}}{2014}]%
        {hill2014nl}
\bibfield{author}{\bibinfo{person}{Emily Hill}, \bibinfo{person}{Manuel
  Roldan-Vega}, \bibinfo{person}{Jerry~Alan Fails}, {and} \bibinfo{person}{Greg
  Mallet}.} \bibinfo{year}{2014}\natexlab{}.
\newblock \showarticletitle{NL-based query refinement and contextualized code
  search results: A user study}. In \bibinfo{booktitle}{\emph{Software
  Maintenance, Reengineering and Reverse Engineering (CSMR-WCRE), 2014 Software
  Evolution Week-IEEE Conference on}}. IEEE, \bibinfo{pages}{34--43}.
\newblock


\bibitem[\protect\citeauthoryear{Hochreiter and Schmidhuber}{Hochreiter and
  Schmidhuber}{1997}]%
        {hochreiter1997long}
\bibfield{author}{\bibinfo{person}{Sepp Hochreiter} {and}
  \bibinfo{person}{J{\"u}rgen Schmidhuber}.} \bibinfo{year}{1997}\natexlab{}.
\newblock \showarticletitle{Long short-term memory}.
\newblock \bibinfo{journal}{\emph{Neural computation}} \bibinfo{volume}{9},
  \bibinfo{number}{8} (\bibinfo{year}{1997}), \bibinfo{pages}{1735--1780}.
\newblock


\bibitem[\protect\citeauthoryear{Hu, Li, Xia, Lo, and Jin}{Hu
  et~al\mbox{.}}{2018a}]%
        {xing2018DeepCom}
\bibfield{author}{\bibinfo{person}{Xing Hu}, \bibinfo{person}{Ge Li},
  \bibinfo{person}{Xin Xia}, \bibinfo{person}{David Lo}, {and}
  \bibinfo{person}{Zhi Jin}.} \bibinfo{year}{2018}\natexlab{a}.
\newblock \showarticletitle{Deep Code Comment Generation}. In
  \bibinfo{booktitle}{\emph{Proceedings of the 2017 26th IEEE/ACM International
  Conference on Program Comprehension}}. ACM.
\newblock


\bibitem[\protect\citeauthoryear{Hu, Li, Xia, Lo, Lu, and Jin}{Hu
  et~al\mbox{.}}{2018b}]%
        {ijcai2018-314}
\bibfield{author}{\bibinfo{person}{Xing Hu}, \bibinfo{person}{Ge Li},
  \bibinfo{person}{Xin Xia}, \bibinfo{person}{David Lo}, \bibinfo{person}{Shuai
  Lu}, {and} \bibinfo{person}{Zhi Jin}.} \bibinfo{year}{2018}\natexlab{b}.
\newblock \showarticletitle{Summarizing Source Code with Transferred API
  Knowledge}. In \bibinfo{booktitle}{\emph{Proceedings of the Twenty-Seventh
  International Joint Conference on Artificial Intelligence, {IJCAI-18}}}.
  \bibinfo{publisher}{International Joint Conferences on Artificial
  Intelligence Organization}, \bibinfo{pages}{2269--2275}.
\newblock
\urldef\tempurl%
\url{https://doi.org/10.24963/ijcai.2018/314}
\showDOI{\tempurl}


\bibitem[\protect\citeauthoryear{Iyer, Konstas, Cheung, and Zettlemoyer}{Iyer
  et~al\mbox{.}}{2016}]%
        {iyer2016summarizing}
\bibfield{author}{\bibinfo{person}{Srinivasan Iyer}, \bibinfo{person}{Ioannis
  Konstas}, \bibinfo{person}{Alvin Cheung}, {and} \bibinfo{person}{Luke
  Zettlemoyer}.} \bibinfo{year}{2016}\natexlab{}.
\newblock \showarticletitle{Summarizing source code using a neural attention
  model}. In \bibinfo{booktitle}{\emph{Proceedings of the 54th Annual Meeting
  of the Association for Computational Linguistics (Volume 1: Long Papers)}},
  Vol.~\bibinfo{volume}{1}. \bibinfo{pages}{2073--2083}.
\newblock


\bibitem[\protect\citeauthoryear{Jiang, Armaly, and McMillan}{Jiang
  et~al\mbox{.}}{2017}]%
        {jiang2017automatically}
\bibfield{author}{\bibinfo{person}{Siyuan Jiang}, \bibinfo{person}{Ameer
  Armaly}, {and} \bibinfo{person}{Collin McMillan}.}
  \bibinfo{year}{2017}\natexlab{}.
\newblock \showarticletitle{Automatically generating commit messages from diffs
  using neural machine translation}. In \bibinfo{booktitle}{\emph{Proceedings
  of the 32nd IEEE/ACM International Conference on Automated Software
  Engineering}}. IEEE Press, \bibinfo{pages}{135--146}.
\newblock


\bibitem[\protect\citeauthoryear{Keivanloo, Rilling, and Zou}{Keivanloo
  et~al\mbox{.}}{2014}]%
        {keivanloo2014spotting}
\bibfield{author}{\bibinfo{person}{Iman Keivanloo}, \bibinfo{person}{Juergen
  Rilling}, {and} \bibinfo{person}{Ying Zou}.} \bibinfo{year}{2014}\natexlab{}.
\newblock \showarticletitle{Spotting working code examples}. In
  \bibinfo{booktitle}{\emph{Proceedings of the 36th International Conference on
  Software Engineering}}. ACM, \bibinfo{pages}{664--675}.
\newblock


\bibitem[\protect\citeauthoryear{Kim}{Kim}{2014}]%
        {kim2014convolutional}
\bibfield{author}{\bibinfo{person}{Yoon Kim}.} \bibinfo{year}{2014}\natexlab{}.
\newblock \showarticletitle{Convolutional neural networks for sentence
  classification}.
\newblock \bibinfo{journal}{\emph{arXiv preprint arXiv:1408.5882}}
  (\bibinfo{year}{2014}).
\newblock


\bibitem[\protect\citeauthoryear{Kingma and Ba}{Kingma and Ba}{2014}]%
        {kingma2014adam}
\bibfield{author}{\bibinfo{person}{Diederik~P Kingma} {and}
  \bibinfo{person}{Jimmy Ba}.} \bibinfo{year}{2014}\natexlab{}.
\newblock \showarticletitle{Adam: A method for stochastic optimization}.
\newblock \bibinfo{journal}{\emph{arXiv preprint arXiv:1412.6980}}
  (\bibinfo{year}{2014}).
\newblock


\bibitem[\protect\citeauthoryear{Lan and Xu}{Lan and Xu}{2018}]%
        {lan2018neural}
\bibfield{author}{\bibinfo{person}{Wuwei Lan} {and} \bibinfo{person}{Wei Xu}.}
  \bibinfo{year}{2018}\natexlab{}.
\newblock \showarticletitle{Neural Network Models for Paraphrase
  Identification, Semantic Textual Similarity, Natural Language Inference, and
  Question Answering}.
\newblock \bibinfo{journal}{\emph{arXiv preprint arXiv:1806.04330}}
  (\bibinfo{year}{2018}).
\newblock


\bibitem[\protect\citeauthoryear{Li, Monroe, Ritter, Galley, Gao, and
  Jurafsky}{Li et~al\mbox{.}}{2016}]%
        {li2016deep}
\bibfield{author}{\bibinfo{person}{Jiwei Li}, \bibinfo{person}{Will Monroe},
  \bibinfo{person}{Alan Ritter}, \bibinfo{person}{Michel Galley},
  \bibinfo{person}{Jianfeng Gao}, {and} \bibinfo{person}{Dan Jurafsky}.}
  \bibinfo{year}{2016}\natexlab{}.
\newblock \showarticletitle{Deep reinforcement learning for dialogue
  generation}.
\newblock \bibinfo{journal}{\emph{arXiv preprint arXiv:1606.01541}}
  (\bibinfo{year}{2016}).
\newblock


\bibitem[\protect\citeauthoryear{Li, Duan, Zhou, Chu, Ouyang, Wang, and
  Zhou}{Li et~al\mbox{.}}{2018a}]%
        {li2018visual}
\bibfield{author}{\bibinfo{person}{Yikang Li}, \bibinfo{person}{Nan Duan},
  \bibinfo{person}{Bolei Zhou}, \bibinfo{person}{Xiao Chu},
  \bibinfo{person}{Wanli Ouyang}, \bibinfo{person}{Xiaogang Wang}, {and}
  \bibinfo{person}{Ming Zhou}.} \bibinfo{year}{2018}\natexlab{a}.
\newblock \showarticletitle{Visual question generation as dual task of visual
  question answering}. In \bibinfo{booktitle}{\emph{Proceedings of the IEEE
  Conference on Computer Vision and Pattern Recognition}}.
  \bibinfo{pages}{6116--6124}.
\newblock


\bibitem[\protect\citeauthoryear{Li, Jiang, Shang, and Li}{Li
  et~al\mbox{.}}{2018b}]%
        {D18-1421}
\bibfield{author}{\bibinfo{person}{Zichao Li}, \bibinfo{person}{Xin Jiang},
  \bibinfo{person}{Lifeng Shang}, {and} \bibinfo{person}{Hang Li}.}
  \bibinfo{year}{2018}\natexlab{b}.
\newblock \showarticletitle{Paraphrase Generation with Deep Reinforcement
  Learning}. In \bibinfo{booktitle}{\emph{Proceedings of the 2018 Conference on
  Empirical Methods in Natural Language Processing}}.
  \bibinfo{publisher}{Association for Computational Linguistics},
  \bibinfo{pages}{3865--3878}.
\newblock


\bibitem[\protect\citeauthoryear{Liu, Lowe, Serban, Noseworthy, Charlin, and
  Pineau}{Liu et~al\mbox{.}}{2016}]%
        {liu2016not}
\bibfield{author}{\bibinfo{person}{Chia-Wei Liu}, \bibinfo{person}{Ryan Lowe},
  \bibinfo{person}{Iulian~V Serban}, \bibinfo{person}{Michael Noseworthy},
  \bibinfo{person}{Laurent Charlin}, {and} \bibinfo{person}{Joelle Pineau}.}
  \bibinfo{year}{2016}\natexlab{}.
\newblock \showarticletitle{How not to evaluate your dialogue system: An
  empirical study of unsupervised evaluation metrics for dialogue response
  generation}.
\newblock \bibinfo{journal}{\emph{arXiv preprint arXiv:1603.08023}}
  (\bibinfo{year}{2016}).
\newblock


\bibitem[\protect\citeauthoryear{Loyola, Marrese-Taylor, and Matsuo}{Loyola
  et~al\mbox{.}}{2017}]%
        {loyola2017neural}
\bibfield{author}{\bibinfo{person}{Pablo Loyola}, \bibinfo{person}{Edison
  Marrese-Taylor}, {and} \bibinfo{person}{Yutaka Matsuo}.}
  \bibinfo{year}{2017}\natexlab{}.
\newblock \showarticletitle{A neural architecture for generating natural
  language descriptions from source code changes}.
\newblock \bibinfo{journal}{\emph{arXiv preprint arXiv:1704.04856}}
  (\bibinfo{year}{2017}).
\newblock


\bibitem[\protect\citeauthoryear{Lu, Sun, Wang, Lo, and Duan}{Lu
  et~al\mbox{.}}{2015}]%
        {lu2015query}
\bibfield{author}{\bibinfo{person}{Meili Lu}, \bibinfo{person}{Xiaobing Sun},
  \bibinfo{person}{Shaowei Wang}, \bibinfo{person}{David Lo}, {and}
  \bibinfo{person}{Yucong Duan}.} \bibinfo{year}{2015}\natexlab{}.
\newblock \showarticletitle{Query expansion via wordnet for effective code
  search}. In \bibinfo{booktitle}{\emph{Software Analysis, Evolution and
  Reengineering (SANER), 2015 IEEE 22nd International Conference on}}. IEEE,
  \bibinfo{pages}{545--549}.
\newblock


\bibitem[\protect\citeauthoryear{Luong, Pham, and Manning}{Luong
  et~al\mbox{.}}{2015}]%
        {luong2015effective}
\bibfield{author}{\bibinfo{person}{Minh-Thang Luong}, \bibinfo{person}{Hieu
  Pham}, {and} \bibinfo{person}{Christopher~D Manning}.}
  \bibinfo{year}{2015}\natexlab{}.
\newblock \showarticletitle{Effective approaches to attention-based neural
  machine translation}.
\newblock \bibinfo{journal}{\emph{arXiv preprint arXiv:1508.04025}}
  (\bibinfo{year}{2015}).
\newblock


\bibitem[\protect\citeauthoryear{McMillan, Grechanik, Poshyvanyk, Fu, and
  Xie}{McMillan et~al\mbox{.}}{2012}]%
        {mcmillan2012exemplar}
\bibfield{author}{\bibinfo{person}{Collin McMillan}, \bibinfo{person}{Mark
  Grechanik}, \bibinfo{person}{Denys Poshyvanyk}, \bibinfo{person}{Chen Fu},
  {and} \bibinfo{person}{Qing Xie}.} \bibinfo{year}{2012}\natexlab{}.
\newblock \showarticletitle{Exemplar: A source code search engine for finding
  highly relevant applications}.
\newblock \bibinfo{journal}{\emph{IEEE Transactions on Software Engineering}}
  \bibinfo{volume}{38}, \bibinfo{number}{5} (\bibinfo{year}{2012}),
  \bibinfo{pages}{1069--1087}.
\newblock


\bibitem[\protect\citeauthoryear{Medsker and Jain}{Medsker and Jain}{1999}]%
        {medsker1999recurrent}
\bibfield{author}{\bibinfo{person}{Larry Medsker} {and}
  \bibinfo{person}{Lakhmi~C Jain}.} \bibinfo{year}{1999}\natexlab{}.
\newblock \bibinfo{booktitle}{\emph{Recurrent neural networks: design and
  applications}}.
\newblock \bibinfo{publisher}{CRC press}.
\newblock


\bibitem[\protect\citeauthoryear{Mnih, Badia, Mirza, Graves, Lillicrap, Harley,
  Silver, and Kavukcuoglu}{Mnih et~al\mbox{.}}{2016}]%
        {mnih2016asynchronous}
\bibfield{author}{\bibinfo{person}{Volodymyr Mnih},
  \bibinfo{person}{Adria~Puigdomenech Badia}, \bibinfo{person}{Mehdi Mirza},
  \bibinfo{person}{Alex Graves}, \bibinfo{person}{Timothy Lillicrap},
  \bibinfo{person}{Tim Harley}, \bibinfo{person}{David Silver}, {and}
  \bibinfo{person}{Koray Kavukcuoglu}.} \bibinfo{year}{2016}\natexlab{}.
\newblock \showarticletitle{Asynchronous methods for deep reinforcement
  learning}. In \bibinfo{booktitle}{\emph{International conference on machine
  learning}}. \bibinfo{pages}{1928--1937}.
\newblock


\bibitem[\protect\citeauthoryear{Moreno, Aponte, Sridhara, Marcus, Pollock, and
  Vijay-Shanker}{Moreno et~al\mbox{.}}{2013}]%
        {moreno2013automatic}
\bibfield{author}{\bibinfo{person}{Laura Moreno}, \bibinfo{person}{Jairo
  Aponte}, \bibinfo{person}{Giriprasad Sridhara}, \bibinfo{person}{Andrian
  Marcus}, \bibinfo{person}{Lori Pollock}, {and} \bibinfo{person}{K
  Vijay-Shanker}.} \bibinfo{year}{2013}\natexlab{}.
\newblock \showarticletitle{Automatic generation of natural language summaries
  for java classes}. In \bibinfo{booktitle}{\emph{Program Comprehension (ICPC),
  2013 IEEE 21st International Conference on}}. IEEE, \bibinfo{pages}{23--32}.
\newblock


\bibitem[\protect\citeauthoryear{Movshovitz-Attias and Cohen}{Movshovitz-Attias
  and Cohen}{2013}]%
        {movshovitz2013natural}
\bibfield{author}{\bibinfo{person}{Dana Movshovitz-Attias} {and}
  \bibinfo{person}{William~W Cohen}.} \bibinfo{year}{2013}\natexlab{}.
\newblock \showarticletitle{Natural language models for predicting programming
  comments}. In \bibinfo{booktitle}{\emph{Proceedings of the 51st Annual
  Meeting of the Association for Computational Linguistics (Volume 2: Short
  Papers)}}, Vol.~\bibinfo{volume}{2}. \bibinfo{pages}{35--40}.
\newblock


\bibitem[\protect\citeauthoryear{Nguyen, Daum{\'e}~III, and Boyd-Graber}{Nguyen
  et~al\mbox{.}}{2017}]%
        {nguyen2017reinforcement}
\bibfield{author}{\bibinfo{person}{Khanh Nguyen}, \bibinfo{person}{Hal
  Daum{\'e}~III}, {and} \bibinfo{person}{Jordan Boyd-Graber}.}
  \bibinfo{year}{2017}\natexlab{}.
\newblock \showarticletitle{Reinforcement Learning for Bandit Neural Machine
  Translation with Simulated Human Feedback}. In
  \bibinfo{booktitle}{\emph{Proceedings of the 2017 Conference on Empirical
  Methods in Natural Language Processing}}. \bibinfo{pages}{1464--1474}.
\newblock


\bibitem[\protect\citeauthoryear{Novikova, Du{\v{s}}ek, Curry, and
  Rieser}{Novikova et~al\mbox{.}}{2017}]%
        {novikova2017we}
\bibfield{author}{\bibinfo{person}{Jekaterina Novikova},
  \bibinfo{person}{Ond{\v{r}}ej Du{\v{s}}ek}, \bibinfo{person}{Amanda~Cercas
  Curry}, {and} \bibinfo{person}{Verena Rieser}.}
  \bibinfo{year}{2017}\natexlab{}.
\newblock \showarticletitle{Why we need new evaluation metrics for nlg}.
\newblock \bibinfo{journal}{\emph{arXiv preprint arXiv:1707.06875}}
  (\bibinfo{year}{2017}).
\newblock


\bibitem[\protect\citeauthoryear{Papineni, Roukos, Ward, and Zhu}{Papineni
  et~al\mbox{.}}{2002}]%
        {papineni2002bleu}
\bibfield{author}{\bibinfo{person}{Kishore Papineni}, \bibinfo{person}{Salim
  Roukos}, \bibinfo{person}{Todd Ward}, {and} \bibinfo{person}{Wei-Jing Zhu}.}
  \bibinfo{year}{2002}\natexlab{}.
\newblock \showarticletitle{BLEU: a method for automatic evaluation of machine
  translation}. In \bibinfo{booktitle}{\emph{Proceedings of the 40th annual
  meeting on association for computational linguistics}}. Association for
  Computational Linguistics, \bibinfo{pages}{311--318}.
\newblock


\bibitem[\protect\citeauthoryear{Paszke, Gross, Chintala, Chanan, Yang, DeVito,
  Lin, Desmaison, Antiga, and Lerer}{Paszke et~al\mbox{.}}{2017}]%
        {paszke2017automatic}
\bibfield{author}{\bibinfo{person}{Adam Paszke}, \bibinfo{person}{Sam Gross},
  \bibinfo{person}{Soumith Chintala}, \bibinfo{person}{Gregory Chanan},
  \bibinfo{person}{Edward Yang}, \bibinfo{person}{Zachary DeVito},
  \bibinfo{person}{Zeming Lin}, \bibinfo{person}{Alban Desmaison},
  \bibinfo{person}{Luca Antiga}, {and} \bibinfo{person}{Adam Lerer}.}
  \bibinfo{year}{2017}\natexlab{}.
\newblock \showarticletitle{Automatic differentiation in PyTorch}.
\newblock  (\bibinfo{year}{2017}).
\newblock


\bibitem[\protect\citeauthoryear{Paulus, Xiong, and Socher}{Paulus
  et~al\mbox{.}}{2017}]%
        {paulus2017deep}
\bibfield{author}{\bibinfo{person}{Romain Paulus}, \bibinfo{person}{Caiming
  Xiong}, {and} \bibinfo{person}{Richard Socher}.}
  \bibinfo{year}{2017}\natexlab{}.
\newblock \showarticletitle{A deep reinforced model for abstractive
  summarization}.
\newblock \bibinfo{journal}{\emph{arXiv preprint arXiv:1705.04304}}
  (\bibinfo{year}{2017}).
\newblock


\bibitem[\protect\citeauthoryear{Ranzato, Chopra, Auli, and Zaremba}{Ranzato
  et~al\mbox{.}}{2015}]%
        {ranzato2015sequence}
\bibfield{author}{\bibinfo{person}{Marc'Aurelio Ranzato},
  \bibinfo{person}{Sumit Chopra}, \bibinfo{person}{Michael Auli}, {and}
  \bibinfo{person}{Wojciech Zaremba}.} \bibinfo{year}{2015}\natexlab{}.
\newblock \showarticletitle{Sequence level training with recurrent neural
  networks}.
\newblock \bibinfo{journal}{\emph{arXiv preprint arXiv:1511.06732}}
  (\bibinfo{year}{2015}).
\newblock


\bibitem[\protect\citeauthoryear{Rennie, Marcheret, Mroueh, Ross, and
  Goel}{Rennie et~al\mbox{.}}{2017}]%
        {rennie2017self}
\bibfield{author}{\bibinfo{person}{Steven~J Rennie}, \bibinfo{person}{Etienne
  Marcheret}, \bibinfo{person}{Youssef Mroueh}, \bibinfo{person}{Jarret Ross},
  {and} \bibinfo{person}{Vaibhava Goel}.} \bibinfo{year}{2017}\natexlab{}.
\newblock \showarticletitle{Self-critical sequence training for image
  captioning}. In \bibinfo{booktitle}{\emph{CVPR}}, Vol.~\bibinfo{volume}{1}.
  \bibinfo{pages}{3}.
\newblock


\bibitem[\protect\citeauthoryear{Rosasco, Vito, Caponnetto, Piana, and
  Verri}{Rosasco et~al\mbox{.}}{2004}]%
        {rosasco2004loss}
\bibfield{author}{\bibinfo{person}{Lorenzo Rosasco},
  \bibinfo{person}{Ernesto~De Vito}, \bibinfo{person}{Andrea Caponnetto},
  \bibinfo{person}{Michele Piana}, {and} \bibinfo{person}{Alessandro Verri}.}
  \bibinfo{year}{2004}\natexlab{}.
\newblock \showarticletitle{Are loss functions all the same?}
\newblock \bibinfo{journal}{\emph{Neural Computation}} \bibinfo{volume}{16},
  \bibinfo{number}{5} (\bibinfo{year}{2004}), \bibinfo{pages}{1063--1076}.
\newblock


\bibitem[\protect\citeauthoryear{Scholer, Williams, and Turpin}{Scholer
  et~al\mbox{.}}{2004}]%
        {scholer2004query}
\bibfield{author}{\bibinfo{person}{Falk Scholer}, \bibinfo{person}{Hugh~E
  Williams}, {and} \bibinfo{person}{Andrew Turpin}.}
  \bibinfo{year}{2004}\natexlab{}.
\newblock \showarticletitle{Query association surrogates for web search}.
\newblock \bibinfo{journal}{\emph{Journal of the American Society for
  Information Science and Technology}} \bibinfo{volume}{55},
  \bibinfo{number}{7} (\bibinfo{year}{2004}), \bibinfo{pages}{637--650}.
\newblock


\bibitem[\protect\citeauthoryear{Schuster and Paliwal}{Schuster and
  Paliwal}{1997}]%
        {schuster1997bidirectional}
\bibfield{author}{\bibinfo{person}{Mike Schuster} {and}
  \bibinfo{person}{Kuldip~K Paliwal}.} \bibinfo{year}{1997}\natexlab{}.
\newblock \showarticletitle{Bidirectional recurrent neural networks}.
\newblock \bibinfo{journal}{\emph{IEEE Transactions on Signal Processing}}
  \bibinfo{volume}{45}, \bibinfo{number}{11} (\bibinfo{year}{1997}),
  \bibinfo{pages}{2673--2681}.
\newblock


\bibitem[\protect\citeauthoryear{Sridhara, Hill, Muppaneni, Pollock, and
  Vijay-Shanker}{Sridhara et~al\mbox{.}}{2010}]%
        {sridhara2010towards}
\bibfield{author}{\bibinfo{person}{Giriprasad Sridhara}, \bibinfo{person}{Emily
  Hill}, \bibinfo{person}{Divya Muppaneni}, \bibinfo{person}{Lori Pollock},
  {and} \bibinfo{person}{K Vijay-Shanker}.} \bibinfo{year}{2010}\natexlab{}.
\newblock \showarticletitle{Towards automatically generating summary comments
  for java methods}. In \bibinfo{booktitle}{\emph{Proceedings of the IEEE/ACM
  international conference on Automated software engineering}}. ACM,
  \bibinfo{pages}{43--52}.
\newblock


\bibitem[\protect\citeauthoryear{Srivastava, Hinton, Krizhevsky, Sutskever, and
  Salakhutdinov}{Srivastava et~al\mbox{.}}{2014}]%
        {srivastava2014dropout}
\bibfield{author}{\bibinfo{person}{Nitish Srivastava},
  \bibinfo{person}{Geoffrey Hinton}, \bibinfo{person}{Alex Krizhevsky},
  \bibinfo{person}{Ilya Sutskever}, {and} \bibinfo{person}{Ruslan
  Salakhutdinov}.} \bibinfo{year}{2014}\natexlab{}.
\newblock \showarticletitle{Dropout: A simple way to prevent neural networks
  from overfitting}.
\newblock \bibinfo{journal}{\emph{The Journal of Machine Learning Research}}
  \bibinfo{volume}{15}, \bibinfo{number}{1} (\bibinfo{year}{2014}),
  \bibinfo{pages}{1929--1958}.
\newblock


\bibitem[\protect\citeauthoryear{{Stack Overflow}}{{Stack Overflow}}{2018}]%
        {StackOverflow}
\bibfield{author}{\bibinfo{person}{{Stack Overflow}}.}
  \bibinfo{year}{2018}\natexlab{}.
\newblock \bibinfo{title}{{Stack Overflow}}.
\newblock \bibinfo{howpublished}{\url{https://stackoverflow.com/}}.
\newblock


\bibitem[\protect\citeauthoryear{Sutskever, Vinyals, and Le}{Sutskever
  et~al\mbox{.}}{2014}]%
        {sutskever2014sequence}
\bibfield{author}{\bibinfo{person}{Ilya Sutskever}, \bibinfo{person}{Oriol
  Vinyals}, {and} \bibinfo{person}{Quoc~V Le}.}
  \bibinfo{year}{2014}\natexlab{}.
\newblock \showarticletitle{Sequence to sequence learning with neural
  networks}. In \bibinfo{booktitle}{\emph{Advances in neural information
  processing systems}}. \bibinfo{pages}{3104--3112}.
\newblock


\bibitem[\protect\citeauthoryear{Sutton and Barto}{Sutton and Barto}{1998}]%
        {sutton1998introduction}
\bibfield{author}{\bibinfo{person}{Richard~S Sutton} {and}
  \bibinfo{person}{Andrew~G Barto}.} \bibinfo{year}{1998}\natexlab{}.
\newblock \bibinfo{booktitle}{\emph{Introduction to reinforcement learning}}.
  Vol.~\bibinfo{volume}{135}.
\newblock \bibinfo{publisher}{MIT press Cambridge}.
\newblock


\bibitem[\protect\citeauthoryear{Sutton, McAllester, Singh, and Mansour}{Sutton
  et~al\mbox{.}}{2000}]%
        {sutton2000policy}
\bibfield{author}{\bibinfo{person}{Richard~S Sutton}, \bibinfo{person}{David~A
  McAllester}, \bibinfo{person}{Satinder~P Singh}, {and}
  \bibinfo{person}{Yishay Mansour}.} \bibinfo{year}{2000}\natexlab{}.
\newblock \showarticletitle{Policy gradient methods for reinforcement learning
  with function approximation}. In \bibinfo{booktitle}{\emph{Advances in neural
  information processing systems}}. \bibinfo{pages}{1057--1063}.
\newblock


\bibitem[\protect\citeauthoryear{Tang, Duan, Qin, Yan, and Zhou}{Tang
  et~al\mbox{.}}{2017}]%
        {tang2017question}
\bibfield{author}{\bibinfo{person}{Duyu Tang}, \bibinfo{person}{Nan Duan},
  \bibinfo{person}{Tao Qin}, \bibinfo{person}{Zhao Yan}, {and}
  \bibinfo{person}{Ming Zhou}.} \bibinfo{year}{2017}\natexlab{}.
\newblock \showarticletitle{Question answering and question generation as dual
  tasks}.
\newblock \bibinfo{journal}{\emph{arXiv preprint arXiv:1706.02027}}
  (\bibinfo{year}{2017}).
\newblock


\bibitem[\protect\citeauthoryear{Vinayakarao, Sarma, Purandare, Jain, and
  Jain}{Vinayakarao et~al\mbox{.}}{2017}]%
        {Vinayakarao:2017:AIS:3018661.3018691}
\bibfield{author}{\bibinfo{person}{Venkatesh Vinayakarao},
  \bibinfo{person}{Anita Sarma}, \bibinfo{person}{Rahul Purandare},
  \bibinfo{person}{Shuktika Jain}, {and} \bibinfo{person}{Saumya Jain}.}
  \bibinfo{year}{2017}\natexlab{}.
\newblock \showarticletitle{ANNE: Improving Source Code Search Using Entity
  Retrieval Approach}. In \bibinfo{booktitle}{\emph{Proceedings of the Tenth
  ACM International Conference on Web Search and Data Mining}}
  \emph{(\bibinfo{series}{WSDM '17})}. \bibinfo{publisher}{ACM},
  \bibinfo{address}{New York, NY, USA}, \bibinfo{pages}{211--220}.
\newblock
\showISBNx{978-1-4503-4675-7}
\urldef\tempurl%
\url{https://doi.org/10.1145/3018661.3018691}
\showDOI{\tempurl}


\bibitem[\protect\citeauthoryear{Voorhees et~al\mbox{.}}{Voorhees
  et~al\mbox{.}}{1999}]%
        {voorhees1999trec}
\bibfield{author}{\bibinfo{person}{Ellen~M Voorhees} {et~al\mbox{.}}}
  \bibinfo{year}{1999}\natexlab{}.
\newblock \showarticletitle{The TREC-8 Question Answering Track Report.}. In
  \bibinfo{booktitle}{\emph{Trec}}, Vol.~\bibinfo{volume}{99}.
  \bibinfo{pages}{77--82}.
\newblock


\bibitem[\protect\citeauthoryear{Wan, Zhao, Yang, Xu, Ying, Wu, and Yu}{Wan
  et~al\mbox{.}}{2018}]%
        {wan2018improving}
\bibfield{author}{\bibinfo{person}{Yao Wan}, \bibinfo{person}{Zhou Zhao},
  \bibinfo{person}{Min Yang}, \bibinfo{person}{Guandong Xu},
  \bibinfo{person}{Haochao Ying}, \bibinfo{person}{Jian Wu}, {and}
  \bibinfo{person}{Philip~S Yu}.} \bibinfo{year}{2018}\natexlab{}.
\newblock \showarticletitle{Improving automatic source code summarization via
  deep reinforcement learning}. In \bibinfo{booktitle}{\emph{Proceedings of the
  33rd ACM/IEEE International Conference on Automated Software Engineering}}.
  ACM, \bibinfo{pages}{397--407}.
\newblock


\bibitem[\protect\citeauthoryear{Wang, Yu, Zhang, Gong, Xu, Wang, Zhang, and
  Zhang}{Wang et~al\mbox{.}}{2017}]%
        {wang2017irgan}
\bibfield{author}{\bibinfo{person}{Jun Wang}, \bibinfo{person}{Lantao Yu},
  \bibinfo{person}{Weinan Zhang}, \bibinfo{person}{Yu Gong},
  \bibinfo{person}{Yinghui Xu}, \bibinfo{person}{Benyou Wang},
  \bibinfo{person}{Peng Zhang}, {and} \bibinfo{person}{Dell Zhang}.}
  \bibinfo{year}{2017}\natexlab{}.
\newblock \showarticletitle{Irgan: A minimax game for unifying generative and
  discriminative information retrieval models}. In
  \bibinfo{booktitle}{\emph{Proceedings of the 40th International ACM SIGIR
  conference on Research and Development in Information Retrieval}}. ACM,
  \bibinfo{pages}{515--524}.
\newblock


\bibitem[\protect\citeauthoryear{Wang, Peng, and Zhang}{Wang
  et~al\mbox{.}}{2018}]%
        {wang2018comment}
\bibfield{author}{\bibinfo{person}{Xiaoran Wang}, \bibinfo{person}{Yifan Peng},
  {and} \bibinfo{person}{Benwen Zhang}.} \bibinfo{year}{2018}\natexlab{}.
\newblock \showarticletitle{Comment Generation for Source Code: State of the
  Art, Challenges and Opportunities}.
\newblock \bibinfo{journal}{\emph{arXiv preprint arXiv:1802.02971}}
  (\bibinfo{year}{2018}).
\newblock


\bibitem[\protect\citeauthoryear{{Wikipedia contributors}}{{Wikipedia
  contributors}}{2019}]%
        {wiki:google_img_labeler}
\bibfield{author}{\bibinfo{person}{{Wikipedia contributors}}.}
  \bibinfo{year}{2019}\natexlab{}.
\newblock \bibinfo{title}{Google Image Labeler --- {Wikipedia}{,} The Free
  Encyclopedia}.
\newblock
\newblock
\urldef\tempurl%
\url{https://en.wikipedia.org/w/index.php?title=Google_Image_Labeler&oldid=881738511}
\showURL{%
\tempurl}
\newblock
\shownote{[Online; accessed 4-February-2019].}


\bibitem[\protect\citeauthoryear{Williams}{Williams}{1992}]%
        {williams1992simple}
\bibfield{author}{\bibinfo{person}{Ronald~J Williams}.}
  \bibinfo{year}{1992}\natexlab{}.
\newblock \showarticletitle{Simple statistical gradient-following algorithms
  for connectionist reinforcement learning}.
\newblock \bibinfo{journal}{\emph{Machine learning}} \bibinfo{volume}{8},
  \bibinfo{number}{3-4} (\bibinfo{year}{1992}), \bibinfo{pages}{229--256}.
\newblock


\bibitem[\protect\citeauthoryear{Yao, Weld, Chen, and Sun}{Yao
  et~al\mbox{.}}{2018}]%
        {yao2018staqc}
\bibfield{author}{\bibinfo{person}{Ziyu Yao}, \bibinfo{person}{Daniel~S Weld},
  \bibinfo{person}{Wei-Peng Chen}, {and} \bibinfo{person}{Huan Sun}.}
  \bibinfo{year}{2018}\natexlab{}.
\newblock \showarticletitle{StaQC: A Systematically Mined Question-Code Dataset
  from Stack Overflow}.
\newblock \bibinfo{journal}{\emph{arXiv preprint arXiv:1803.09371}}
  (\bibinfo{year}{2018}).
\newblock


\bibitem[\protect\citeauthoryear{Zaremba, Sutskever, and Vinyals}{Zaremba
  et~al\mbox{.}}{2014}]%
        {zaremba2014recurrent}
\bibfield{author}{\bibinfo{person}{Wojciech Zaremba}, \bibinfo{person}{Ilya
  Sutskever}, {and} \bibinfo{person}{Oriol Vinyals}.}
  \bibinfo{year}{2014}\natexlab{}.
\newblock \showarticletitle{Recurrent neural network regularization}.
\newblock \bibinfo{journal}{\emph{arXiv preprint arXiv:1409.2329}}
  (\bibinfo{year}{2014}).
\newblock


\end{thebibliography}
